\documentclass[useAMS,usenatbib,usegraphicx]{mn2e}

\title[An extremely isolated Local Group globular cluster]{Deep Gemini/GMOS imaging of an 
extremely isolated globular cluster in the Local Group}
\author[A.~D.~Mackey et al.]{A.~D.~Mackey$^{1}$, 
A.~M.~N.~Ferguson$^{1}$, M.~J.~Irwin$^{2}$, N.~F.~Martin$^{3}$, A.~P.~Huxor$^{4}$,\newauthor
N.~R.~Tanvir$^{5}$, S.~C.~Chapman$^{2}$, R.~A.~Ibata$^{6}$, G.~F.~Lewis$^{7}$, 
A.~W.~McConnachie$^{8}$\\
$^{1}$Institute for Astronomy, University of Edinburgh, Royal Observatory, Blackford Hill,
Edinburgh, EH9 3HJ, UK \\
$^{2}$Institute of Astronomy, University of Cambridge, Madingley Road, Cambridge, CB3 0HA, UK \\
$^{3}$Max-Planck-Institut f\"{u}r Astronomie, K\"{o}nigstuhl 17, D-69117 Heidelberg, Germany \\
$^{4}$Department of Physics, University of Bristol, Tyndall Avenue, Bristol, BS8 1TL, UK \\
$^{5}$Department of Physics and Astronomy, University of Leicester, University Road, Leicester,
LE1 7RH, UK \\
$^{6}$Observatoire Astronomique, Universit\'{e} de Strasbourg, CNRS, 11, rue de l'Universit\'{e},
F-67000 Strasbourg, France \\
$^{7}$Institute of Astronomy, School of Physics, A29, University of Sydney, NSW 2006, Australia \\
$^{8}$NRC Herzberg Institute for Astrophysics, 5071 West Saanich Road, Victoria, British Columbia,
Canada V9E 2E7}
\begin{document}

\date{Accepted 2009 September 5. Received 2009 September 1; in original form 2009 June 9.}

\pagerange{\pageref{firstpage}--\pageref{lastpage}} \pubyear{2009}

\maketitle

\label{firstpage}

\begin{abstract}
We report on deep imaging of a remote M31 globular cluster, MGC1, obtained with 
Gemini/GMOS. Our colour-magnitude diagram for this object extends $\sim 5$ magnitudes
below the tip of the red giant branch and exhibits features consistent with an ancient
metal-poor stellar population, including a long, well-populated horizontal branch. 
The red giant branch locus suggests MGC1 has a metal abundance $[$M$/$H$] \approx -2.3$.
We measure the distance to MGC1 and find that it lies $\sim 160$ kpc in front of M31
with a distance modulus $\mu = 23.95 \pm 0.06$. Combined with its large projected 
separation of $R_{{\rm p}} = 117$ kpc from M31 this implies a deprojected radius of 
$R_{{\rm gc}} = 200 \pm 20$ kpc, rendering it the most isolated known globular cluster 
in the Local Group by some considerable margin. We construct a radial brightness profile 
for MGC1 and show that it is both centrally compact and rather luminous, with $M_V = -9.2$.
Remarkably, the cluster profile shows no evidence for a tidal limit and we are able
to trace it to a radius of at least $450$ pc, and possibly as far as $\sim 900$ pc.
The profile exhibits a power-law fall-off with exponent $\gamma = -2.5$, breaking to 
$\gamma = -3.5$ in its outermost parts. This core-halo structure is broadly consistent 
with expectations derived from numerical models, and suggests that MGC1 has spent many 
gigayears in isolation.
\end{abstract}


\section{Introduction}
Cosmological models of structure formation predict that galaxies are built up via the
hierarchical accretion and merger of many smaller subsystems over time. It is expected
that the signatures of these galaxy assembly processes may be seen in the outskirts
of present-day large galaxies where dynamical timescales are very long. Globular 
clusters, as luminous compact objects that are found out to distant radii in the
halos of massive galaxies, are thus potentially extremely useful probes of galaxy 
formation. This is especially the case for systems beyond a few Mpc where globular
clusters are often the only accessible tracers of ancient stellar populations.

In terms of testing and refining these ideas, galaxies in the Local Group play a 
central role. They are close enough that their various components, including individual
globular clusters, may be studied in detail as resolved stellar populations. M31 is
of particular significance as it is the nearest large galaxy to our own, and possesses
the most extensive globular cluster system in the Local Group. This galaxy has been
targeted by a number of wide-field imaging surveys in recent years and these
have revealed a wealth of low-brightness substructure in its outer regions
\citep[e.g.,][]{ibata:01,ferguson:02,ibata:07,mcconnachie:09} as well as many
new members of its halo globular cluster system, extending to very large radii
\citep[e.g.,][]{huxor:05,mackey:06,mackey:07,huxor:08}.

Although the M31 globular cluster system shares many similarities with that of the
Milky Way, as the sample of outer halo M31 globular clusters has grown and been studied
in more detail over the past few years several clear differences have emerged. Among 
these are the existence in the M31 system of very extended yet still rather luminous 
clusters \citep{huxor:05,mackey:06,huxor:08} that do not seem to have matching counterparts
in the Milky Way, and indications that M31 may possess many more very luminous compact 
globular clusters in its outer regions than does the Milky Way 
\citep[e.g.,][]{mackey:07,huxor:09}.
The origin of these distinctions is not yet clear but it has been speculated that they 
may reflect differences in the formation and/or accretion histories of the Milky Way and 
M31. It is therefore of considerable interest to obtain as much additional information as 
possible about the remote globular cluster system in M31.

One of the most intriguing newly-found objects in the M31 halo is a bright globular 
cluster discovered by \citet{martin:06} in survey observations from the MegaCam imager
on the Canada-France-Hawaii Telescope (CFHT). The cluster lies at $\alpha = 00$:$50$:$42.45$,\ 
$\delta = +32$:$54$:$58.70$ (J2000.0), or roughly $\sim 8.5\degr$ from the centre of 
M31\footnote{Assumed to be at $\alpha = 00$:$42$:$44.31$,\ $\delta = +41$:$16$:$09.4$ (J2000.0).}. 
Adopting the standard distance to M31 of $\approx 780$ kpc \citep{mcconnachie:05}, this 
angular separation corresponds to a projected radius of $R_{{\rm p}} = 117$ kpc, placing the 
new globular cluster among most remote such objects yet known in the M31 system and, indeed, 
in the Local Group as a whole. Following the 
nomenclature adopted in the latest version (V3.5) of the Revised Bologna Catalogue (RBC) of 
M31 globular clusters \citep{galleti:07}, we hereafter refer to this object as MGC1. 

\citet{martin:06} used their CFHT/MegaCam discovery imagery to measure a few characteristics 
of the cluster. They found it to be a compact, luminous object with half-light radius
$r_h \sim 2.3 \pm 0.2$ pc and absolute integrated magnitude $M_V = -8.5 \pm 0.3$.
Their CMD showed a well-populated red giant branch, and a probable horizontal branch 
near their faint detection limit. Isochrone fitting implied a metal abundance
$[$Fe$/$H$] \approx -1.3$. Most intriguingly, the observed brightness of the tip of the 
red giant branch, along with that of the putative horizontal branch, suggested that
MGC1 may lie considerably closer than M31, meaning that the cluster may be 
extremely isolated indeed with a deprojected galactocentric radius of $175 \pm 55$ kpc.

Two groups have published radial velocities for MGC1: \citet{galleti:07} obtained
$v_r = -312 \pm 17$ km$\,$s$^{-1}$ from low-resolution spectroscopy, while
\citet{alvesbrito:09} measured $v_r = -354.7 \pm 2.2$ km$\,$s$^{-1}$ from a high
resolution Keck spectrum. These measurements are consistent with the systemic
velocity of M31 ($v_r = -300$ km$\,$s$^{-1}$), suggesting that MGC1 is a bona fide
member of M31's outer halo globular cluster system. In this regard it is of considerable 
significance as very few such objects have yet been discovered -- there are
only three other known globular clusters in M31 with $R_{{\rm p}} \ga 70$ kpc
\citep{mackey:07,huxor:08}. Indeed there are also very few individual halo stars
that have been identified at projected radii comparable to that of MGC1 
\citep[e.g.,][]{kalirai:06,chapman:08}.

In this paper we present results from deep Gemini follow-up imaging of this intriguing 
cluster, with the particular goals of (i) obtaining an accurate distance measurement in 
order to understand properly its position in the M31 system; (ii) obtaining a precise
estimate of its metal abundance; and (iii) making a thorough analysis of its structure.

\section{Observations and data reduction}
\subsection{Imaging and photometry}
\label{ss:imagephot}
Our deep imaging of MGC1 was carried out in queue mode on the night of 2007 October 13 using the 
Gemini Multi-Object Spectrograph (GMOS) at the $8.1$m Gemini North telescope on Mauna Kea, Hawaii. 
The data were obtained in clear, photometric conditions and under excellent seeing ($0.4-0.5\arcsec$).
The GMOS imager \citep{hook:04} is comprised of three adjacent $2048\times4096$ pixel CCDs 
separated by gaps of $\sim 2.8\arcsec$, and has a field of view (which does not cover the full 
CCD package) of $5.5\arcmin \times 5.5\arcmin$. To take advantage of the high quality conditions 
we employed unbinned imaging, resulting in a plate scale of $0.0727$ arcsec$/$pixel. We 
obtained our observations in the GMOS $g\arcmin$, $r\arcmin$, and $i\arcmin$ filters, which are 
very similar to the $g$, $r$, and $i$ filters used by the Sloan Digital Sky Survey (SDSS). Details 
of the SDSS photometric system may be found in \citet{fukugita:96}. We took six images per filter, 
arranged in a $3\times2$ dither pattern with a step size of $5\arcsec$ designed to eliminate the
gaps between the CCDs and provide continuous coverage of our field. Exposure durations were $400$s 
per image for the $r\arcmin$ and $i\arcmin$ filters and $600$s per image for the $g\arcmin$ filter. 

We reduced our data using the GMOS reduction package in {\sc iraf}. Appropriate bias
and flat-field images obtained as part of the standard GMOS baseline calibrations were downloaded 
from the Gemini science archive and then applied to each exposure using the {\sc gireduce} task. 
The three CCD frames in a given exposure were next mosaicked into a single frame using the 
{\sc gmosaic} task, and the six mosaicked frames for a given filter were then combined into
a single image using the {\sc imcoadd} task. 

\begin{figure*}
\begin{minipage}{175mm}
\begin{center}
\includegraphics[width=90mm]{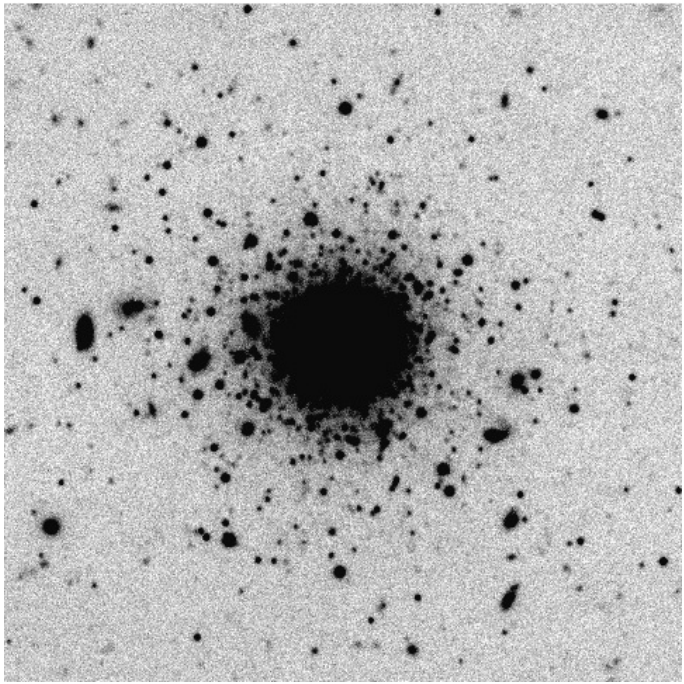}
\caption{Central $70\arcsec \times 70\arcsec$ region of our combined $2400$s $i\arcmin$-band 
GMOS image, showing MGC1. Stellar objects in this image have a FWHM\ $\approx 6.2$ pixels, 
or $0.45\arcsec$. North is towards the top of the image and east is to the left.}
\label{f:cluster}
\end{center}
\end{minipage}
\end{figure*}

Fig. \ref{f:cluster} shows the central $70\arcsec \times 70\arcsec$ region of our combined 
$2400$s $i\arcmin$-band image. Stellar objects in this image have a FWHM\ $\approx 6.2$ pixels, 
or $0.45\arcsec$, and the outer region of MGC1 is clearly resolved into stars. 
The inner part of the cluster is unresolved, but, as described below, the images nonetheless 
contain a sufficient number of individual cluster members to produce a well-populated 
colour-magnitude diagram (CMD). 

We performed photometric measurements on our combined images using the stand-alone 
versions of {\sc daophot ii} and {\sc allstar ii} \citep{stetson:87}. Because of the
crowded nature of the region of interest, full point-spread function (PSF) fitting photometry 
was necessary. For each combined image we used {\sc daophot ii} to conduct a first pass of 
object detection, and selected $\sim 75$ relatively bright, isolated stars to construct an 
initial PSF. Some experimentation showed that the best fit was obtained by selecting a Gaussian 
PSF model and allowing quadratic variation across the field of view. Stars with an error in 
their fit more than three times the average were removed from the list, and the PSF redefined. 
After iterating until convergence, at which point there were typically still $\sim 50-60$ stars 
defining the PSF model, we used {\sc allstar} to subtract from the image all stars in close 
proximity to those used to construct the PSF. The now-isolated PSF stars on this subtracted image 
were used to recalculate and further refine the PSF model. Next, we used {\sc allstar} to apply 
this model to the original image and subtract all known stars. This subtracted image was then 
run through {\sc daophot} in order to find faint objects missed in the first detection pass. 
We then took the original image and our final PSF model and used {\sc allstar} to perform 
photometric measurements on the complete list of detected objects.

In order to eliminate non-stellar objects or objects with poor photometry, we passed 
the resulting list of measurements through several quality filters based on information
calculated during the PSF fitting. Specifically, we filtered objects by the $\chi^2$ of
their fit, their estimated photometric error, and their measured sharpness relative to the 
PSF model. To define suitable ranges in these parameters for well-measured point sources 
we selected all objects in an annulus of radius $12\arcsec-40\arcsec$ centered on the cluster, 
corresponding to the region populated almost exclusively by cleanly resolved member stars. 
These objects defined clear, narrow loci in terms of $\chi^2$, photometric error and sharpness 
as functions of magnitude in each of the three filters, which we used to filter the final 
list of detections across the full field of view. 

\subsection{Photometric calibration}
\label{ss:photcal}
In addition to observing our target field we also obtained imaging of photometric standard 
stars in one equatorial field, around star SA 92-249 \citep{landolt:92}. This imaging was
conducted through all three filters immediately after the observations of our primary field 
were complete. We took two exposures per filter, with durations of $4$s, $3$s, and $2$s each 
in $g\arcmin$, $r\arcmin$, and $i\arcmin$, respectively. An identical reduction process to
that followed for the primary imaging was applied to the observations of our photometric
standards to produce one mosaicked, combined image per filter. We then used the stand-alone
version of {\sc daophot ii} to perform aperture photometry on the standard stars in these 
combined images, using an aperture equal in radius to that for the PSF models we used to
obtain photometric measurements from the main cluster images. 

In all, six photometric standard stars were covered by the GMOS field of view. Unfortunately,
however, we could only obtain accurate photometric measurements for four of these -- one star 
was completely vignetted by the GMOS On-Instrument Wavefront Sensor (OIWFS), which facilitates 
precise guiding and use of the telescope's active optics system, while a second star fell on 
the gap between the CCDs in half the images. 

Our standard field lies within the SDSS footprint; however the stars are sufficiently bright
that they are flagged as saturated in the SDSS catalogue. We therefore employed an alternative
method of obtaining their SDSS $g$-, $r$-, and $i$-band AB-magnitudes, by transforming the
\citet{landolt:92} Johnson-Kron-Cousins $BVRI$ measurements using the relations listed on 
the SDSS Data Release 7 (DR7) 
website\footnote{{\scriptsize {\it http://www.sdss.org/dr7/algorithms/sdssUBVRITransform.html}}}. 
Specifically, we used inverted versions of Lupton's $ugriz$ to $BVRI$ transformations:
\begin{eqnarray}
g &=& 0.6489\,B + 0.3511\,I - 0.1460 \nonumber \\
r &=& 1.3088\,R - 0.3088\,I + 0.0704 \\
i &=& 0.2570\,R + 0.7430\,I + 0.3208 \nonumber
\end{eqnarray}
The other alternative transformations listed on the DR7 webpage, in particular 
those of \citet{jester:05}; \citet*{bilir:05}; and \citet*{jordi:06} give matching $gri$ 
magnitudes for our standards to within a few$\,\times\,0.01$ mag. Therefore, the 
selection of any of these sets of relations would not significantly alter our derived 
photometric calibration.

\begin{figure}
\centering
\includegraphics[width=0.23\textwidth]{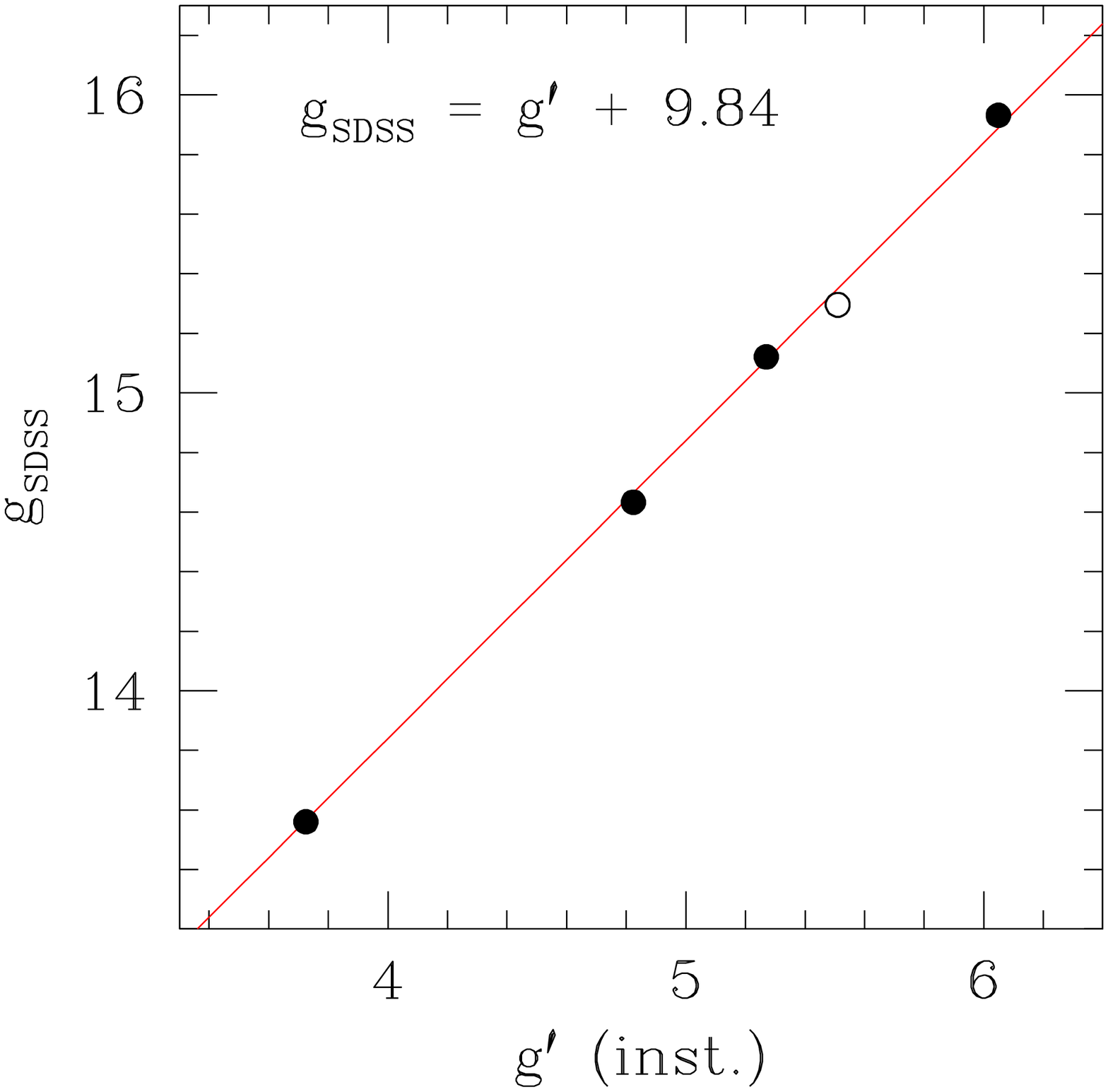}
\hspace{-0.5mm}
\includegraphics[width=0.23\textwidth]{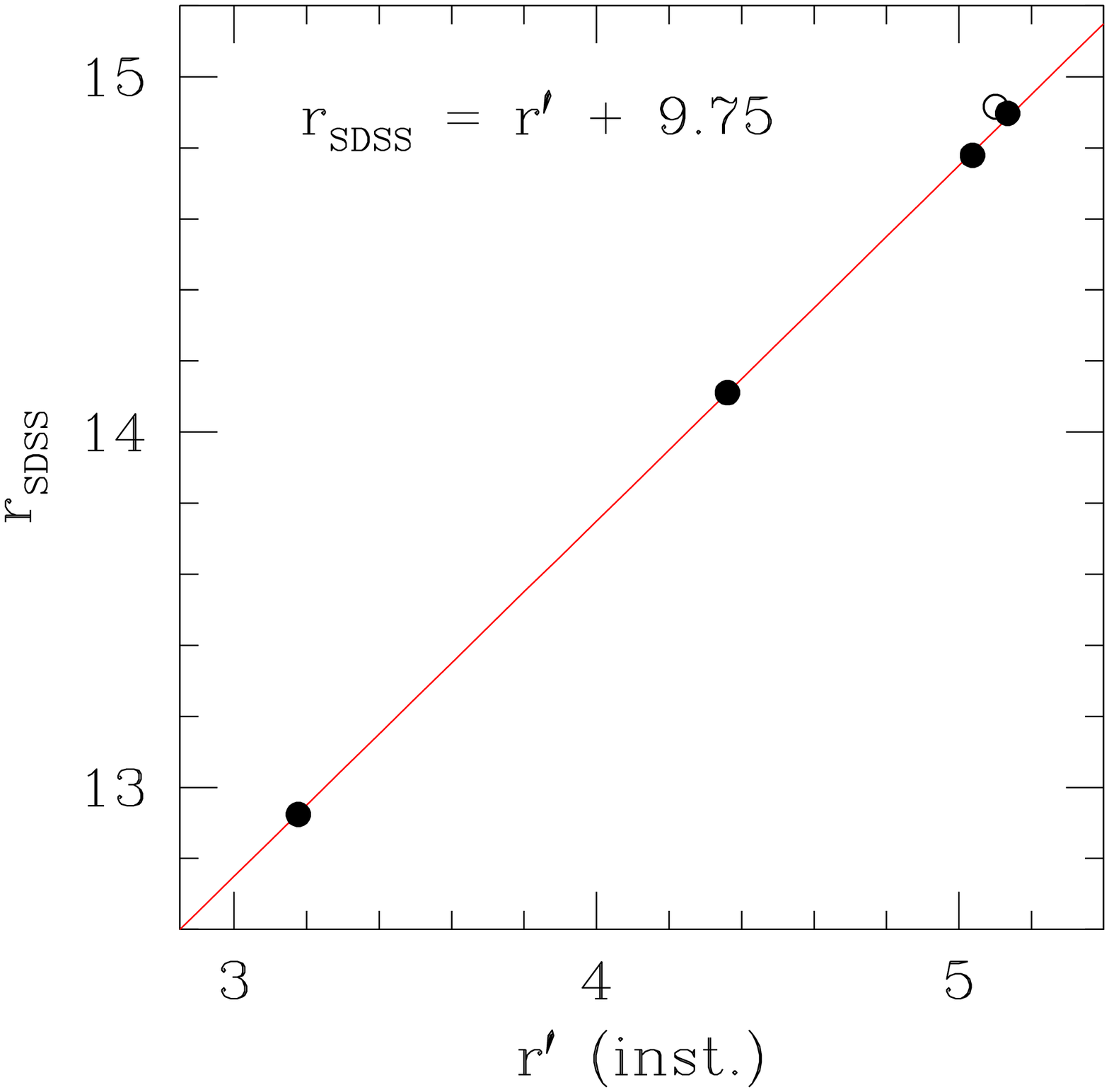}\\
\vspace{1mm}
\includegraphics[width=0.23\textwidth]{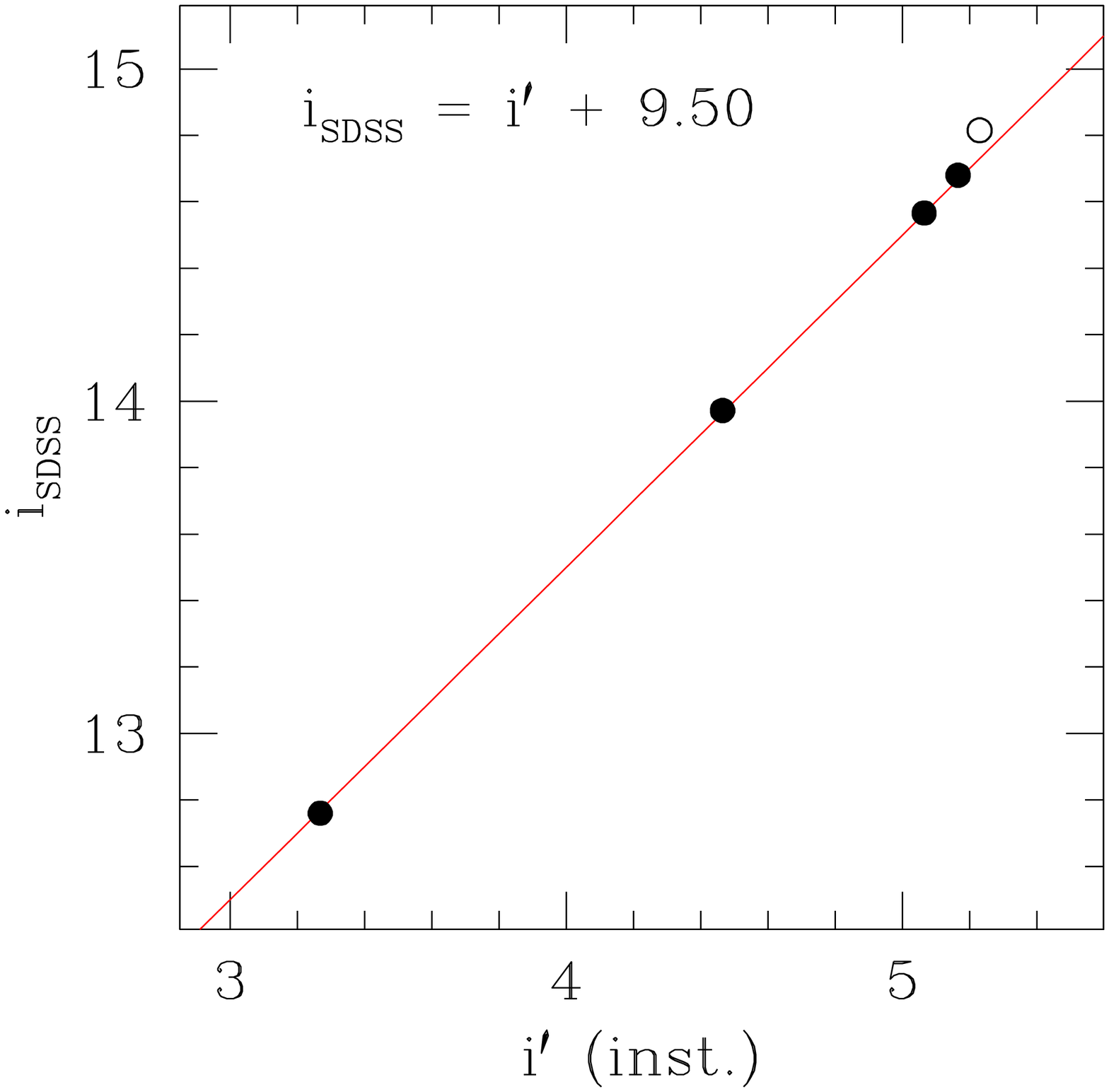}
\caption{Transformations from GMOS instrumental magnitudes to SDSS AB magnitudes derived from 
the observed photometric standards. The four stars with complete measurements are marked as solid 
points, while the star which fell on the CCD gap in half the images is an open point.}
\label{f:inst2sdss}
\end{figure}

Fig. \ref{f:inst2sdss} shows the relationship between our measured GMOS instrumental 
$g\arcmin$, $r\arcmin$ and $i\arcmin$ magnitudes for the standard stars, and their derived 
SDSS $gri$ AB-magnitudes. The instrumental magnitudes as plotted have been corrected to 
the exposure times of the three respective combined cluster images. In each panel the star 
marked with an open circle was that which fell on the gap between the CCDs in half the images. 
We did not use our measurements for this star when calculating the best-fit transformations. 
From Fig. \ref{f:inst2sdss} the relationships between $g\arcmin r\arcmin i\arcmin$ and $gri$ 
are well described by a simple zero-point shift. We determined the best-fit calibration via 
linear regression:
\begin{eqnarray}
g &=& g\arcmin + 9.84 \nonumber \\
r &=& r\arcmin + 9.75 \label{e:inst2sdss}\\
i &=& i\arcmin + 9.50 \nonumber
\end{eqnarray}
Formal (random) uncertainties on these zero-point shifts from the linear regression 
are typically $\pm 0.02$ mag. Even so, the above transformations were derived using only 
four stars which do not span a particularly large range in colour ($0.45 \leq (g-i) \leq 1.35$),
meaning that it was impossible for us to determine colour terms for Eq. \ref{e:inst2sdss}.
\citet{jorgensen:09} has demonstrated that small colour terms are expected in
transformations to the SDSS photometric systems for all GMOS-N filters, and has used a
large number of standard star observations to derive these corrections. Unfortunately
her photometric calibration and colour terms are valid only for data obtained prior to 
November 2004 when the main mirror of Gemini North was re-coated with silver in place 
of aluminium, and so we cannot use them directly for our MGC1 observations. None the less, 
we expect them to give some indication of the size of the systematic errors introduced into 
our photometric calibration by not including any colour terms in Eq. \ref{e:inst2sdss}.

Our colour-magnitude diagrams show that stars in MGC1 span $-0.2 \la (g-r) \la 1.4$
and $-0.4 \la (g-i) \la 1.7$ (see Section \ref{ss:cmd}). The appropriate colour-term
corrections to $gri$ magnitudes from \citet{jorgensen:09} are then:
\begin{eqnarray}
\Delta g_1 &=& 0.066 (g-r) - 0.037 \nonumber \\
\Delta g_2 &=& 0.035 (g-i) - 0.029 \nonumber \\
\Delta r_1 &=& 0.027 (g-r) - 0.016 \nonumber \\
\Delta r_2 &=& 0.017 (g-i) - 0.015 \label{e:colourterms} \\
\Delta i_1 &=& 0.039 (g-r) - 0.020 \nonumber \\
\Delta i_2 &=& 0.025 (g-i) - 0.018 \nonumber
\end{eqnarray}
where the two corrections $\Delta m_1$ and $\Delta m_2$ for a given filter may be 
calculated independently. Over the range of colours spanned by cluster members, these 
would correspond to systematic offsets in our SDSS magnitudes of $-0.04 \la \Delta g \la 0.04$, 
$-0.02 \la \Delta r \la 0.02$, and $-0.03 \la \Delta i \la 0.03$, and in our
SDSS colours of $-0.02 \la \Delta(g-r) \la 0.02$ and $-0.01 \la \Delta(g-i) \la 0.01$.
Such uncertainties are comparable in size to the formal random uncertainties on our
zero-points (Eq. \ref{e:inst2sdss}).
As such, we are confident that in what follows our analysis and conclusions are largely 
unaffected by the fact that we did not include colour terms in our photometric 
calibration. For completeness, where appropriate we explicitly make note of the effects 
these small systematic offsets may have on our measurements of MGC1.

\begin{figure*}
\begin{minipage}{175mm}
\begin{center}
\includegraphics[width=60mm]{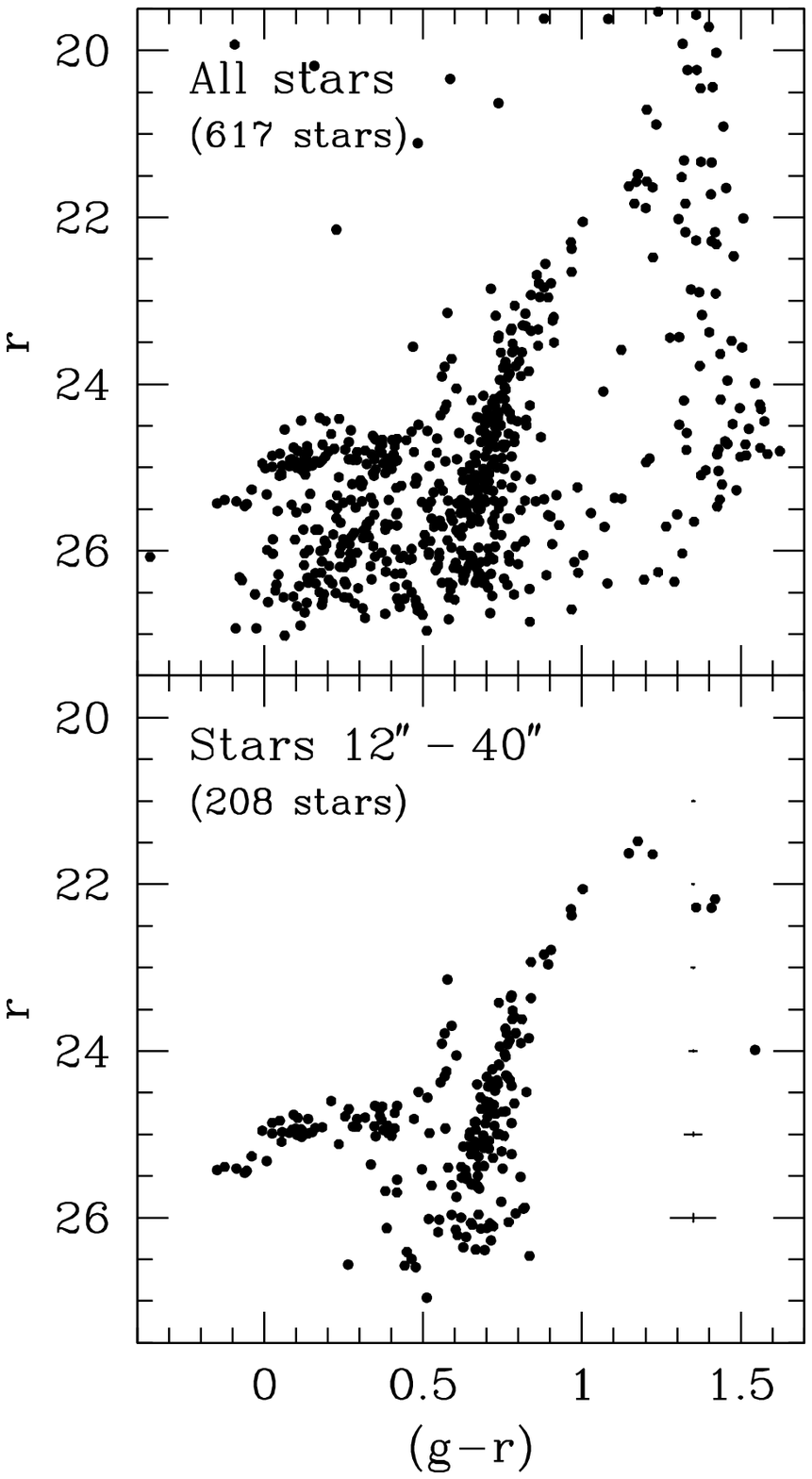}
\hspace{-2mm}
\includegraphics[width=60mm]{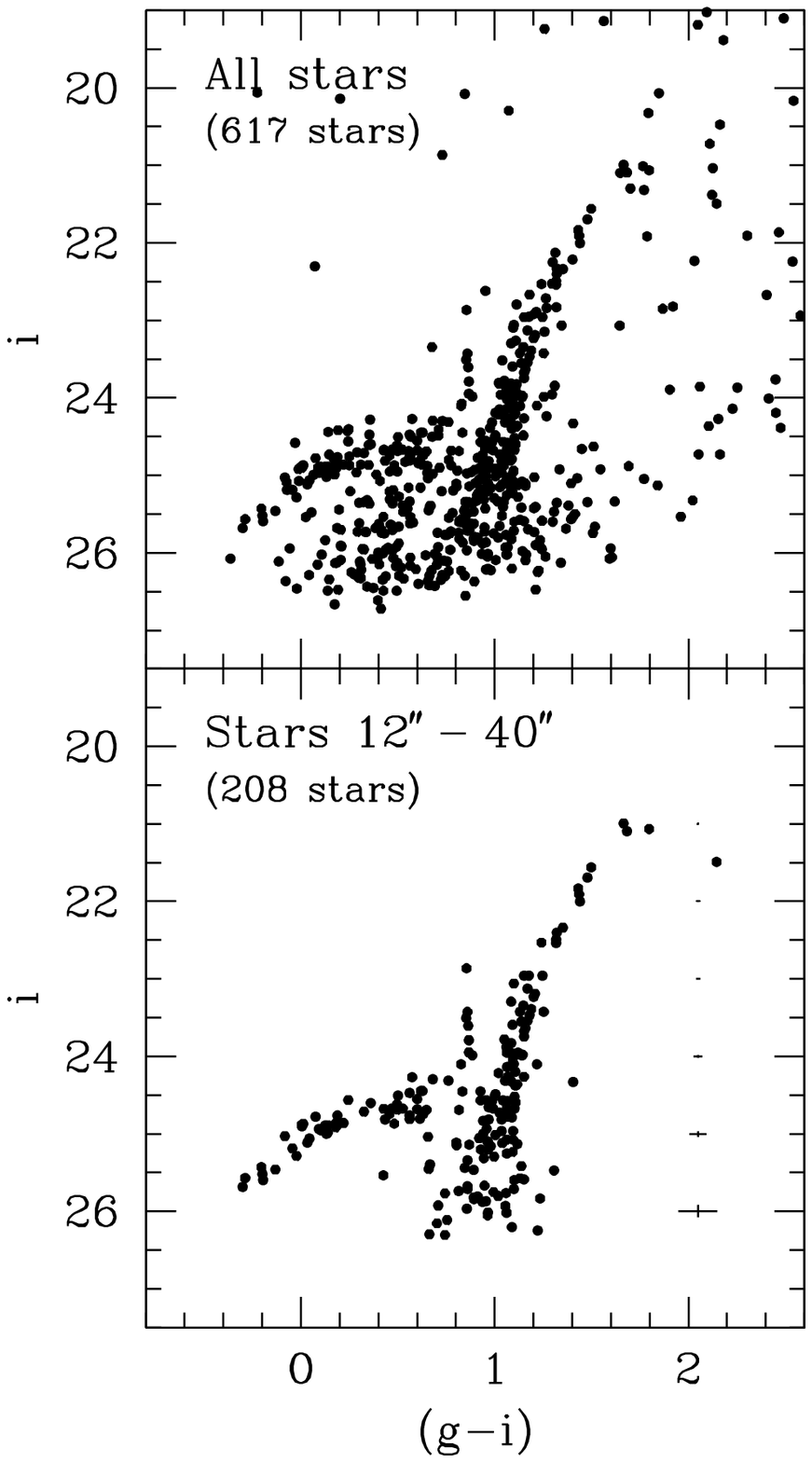}
\caption{Final $(g,r)$ and $(g,i)$ colour-magnitude diagrams for the full GMOS frame 
(upper panels) and for an annulus spanning the radial range $12\arcsec - 40\arcsec$ centred 
on MGC1 (lower panels). Typical photometric uncertainties are marked in the lower panels.}
\label{f:cmds}
\end{center}
\end{minipage}
\end{figure*}

\section{Analysis of the CMD}
\subsection{General properties}
\label{ss:cmd}
Our final $(g,r)$ and $(g,i)$ colour-magnitude diagrams are presented in Fig. \ref{f:cmds}. 
The upper panels show all detected objects which passed successfully through 
the quality filter we described in Section \ref{ss:imagephot}, while the lower panels show
the subset of those objects which lie in the projected radial range $12\arcsec - 40\arcsec$ 
from the centre of MGC1. We determined these radial limits empirically, to exhibit the best-defined 
cluster sequences with minimal degradation due to crowding or background contamination. The lower 
panels also show the typical photometric uncertainties as functions of magnitude.

The cluster sequences are clearly visible in the upper panels in Fig. \ref{f:cmds} as are
several contaminating populations. To the red side of the CMDs, at $(g - r) \ga 1.2$ and 
$(g - i) \ga 1.6$, lies the sequence due to Galactic disk dwarfs; this is most 
prominent in the $(g,r)$ CMD. Also evident are a few members of the Galactic halo 
(e.g., at $(g - i) \la 1.0$ and $i \la 22.5$) and a significant number of faint unresolved 
galaxies (which have $0.0 \la (g - i) \la 1.4$ and $i \ga 25.0$). In the lower panels, 
which cover only a limited area on
the sky, objects from these populations are almost completely absent.

The cluster itself possesses a well-populated steep red-giant branch (RGB) and a horizontal
branch (HB) that extends far to the blue. Our photometry reaches $\sim 1.5$ magnitudes 
below the HB, meaning that this feature is clearly isolated on the CMD. There are no large 
gaps along the measured length of the HB, implying that the cluster is likely to possess a 
population of RR Lyrae stars. Our observational baseline, however, does not span a sufficiently 
long interval to allow a useful variability search to be performed. Overall, the morphology 
of the HB and the implied presence of RR Lyrae stars, together with the steep 
RGB, is indicative of a very old ($\ga 10$ Gyr) metal-poor stellar population.

For ancient systems such as this, the 
locus of the RGB above the level of the HB is mostly dependent on the metal abundance and 
only weakly sensitive to age. The observed shape of this feature therefore provides a 
useful and well-established probe of the cluster metallicity \citep[e.g.,][]{sarajedini:94}.
Meanwhile, the observed level of the HB and the apparent colour of the RGB at the HB level
allow estimates to be made of the distance to the cluster and the amount of foreground
reddening. In order to obtain such measurements for MGC1, we compared our CMDs with
fiducial sequences observed for several Galactic globular clusters in the SDSS photometric
system. We subsequently checked the consistency of our results by fitting
theoretical isochrones from stellar models calculated by two different groups.

\subsection{Comparison with Galactic globular clusters}
\label{ss:ggc}
\citet{an:08} provide CMDs in the SDSS photometric system for $17$
Galactic globular clusters spanning a wide range of metal abundances, 
$-2.3 \la [$Fe$/$H$] \la -0.7$. Unfortunately for the purposes of estimating a photometric 
metallicity for MGC1, many of the CMDs they provide (and in particular those for 
all but one of the clusters with $[$Fe$/$H$] \la -2$) are truncated only a magnitude or two above 
the HB level due to the saturation limits of the SDSS imagery (see their Fig. 12). None the less,
there are two well-populated clusters with abundances in the range of interest that are 
sufficiently distant for the entire extent of their RGBs to be sampled. These are NGC 2419 
and NGC 7006, which have $[$Fe$/$H$] = -2.12$ and $-1.63$, respectively, according to 
\citet{harris:96}.

\begin{figure}
\centering
\includegraphics[width=0.48\textwidth]{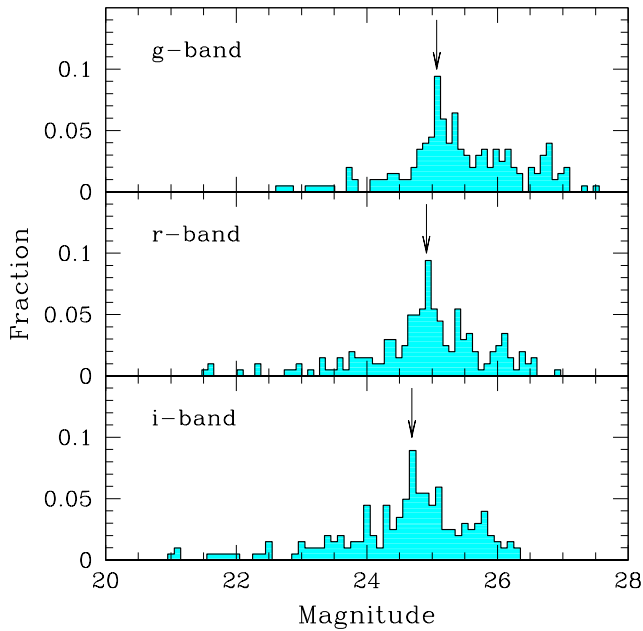}
\caption{Cluster luminosity functions in each passband, constructed using
only stars in the radial range $12\arcsec - 40\arcsec$. The HB level is clearly visible and
is marked with an arrow in each panel.}
\label{f:lumfunc}
\end{figure}

To correctly align the fiducial sequences for these two clusters, we shifted them vertically on 
the CMD so that their HB levels matched that observed for MGC1, and horizontally so
that their RGB colours at the level of the HB matched the observed value for MGC1. 
We determined the appropriate HB level for MGC1 in each of the three passbands 
by plotting luminosity functions. These are presented in Fig. \ref{f:lumfunc} -- the HB 
level is clearly visible in each of the histograms, at $g = 25.1$, $r = 24.9$, and $i = 24.7$. 
The uncertainty in these values is $\pm 0.05$ mag. Note that \citet{an:08} do not provide
fiducial HB sequences for their clusters; however their photometry is publicly available
and these ridgelines are straightforward to determine.

Because \citet{an:08} obtained their fiducial sequences from a homogeneous data set via a 
uniform reduction and photometry pipeline, we were able to impose an additional constraint
in our alignment procedure -- that the implied offsets in distance modulus ($\mu$) and 
foreground reddening between MGC1 and a given reference cluster should be equal on both 
the $(g,i)$ and $(g,r)$ CMDs. That is, for a given reference cluster we tested all likely 
combinations of $\Delta \mu$ and $\Delta E(B-V)$, converted these to vertical and horizontal 
offsets in the observational plane, and determined which combination gave the closest alignment 
between the observed globular cluster fiducial and MGC1 on both the $(g,i)$ and $(g,r)$ CMDs 
simultaneously. To convert $\Delta \mu$ and $\Delta E(B-V)$ to vertical and horizontal 
shifts we used the coefficients from \citet{stoughton:02}, which relate the foreground 
extinction to $E(B-V)$ in SDSS passbands: $A_g = 3.793E(B-V)$, $A_r = 2.751E(B-V)$ and 
$A_i = 2.086E(B-V)$. Given these, it is trivial to show that $E(g-i) = 1.707E(B-V)$ and 
$E(g-r) = 1.042E(B-V)$.

\begin{figure*}
\centering
\includegraphics[width=0.32\textwidth]{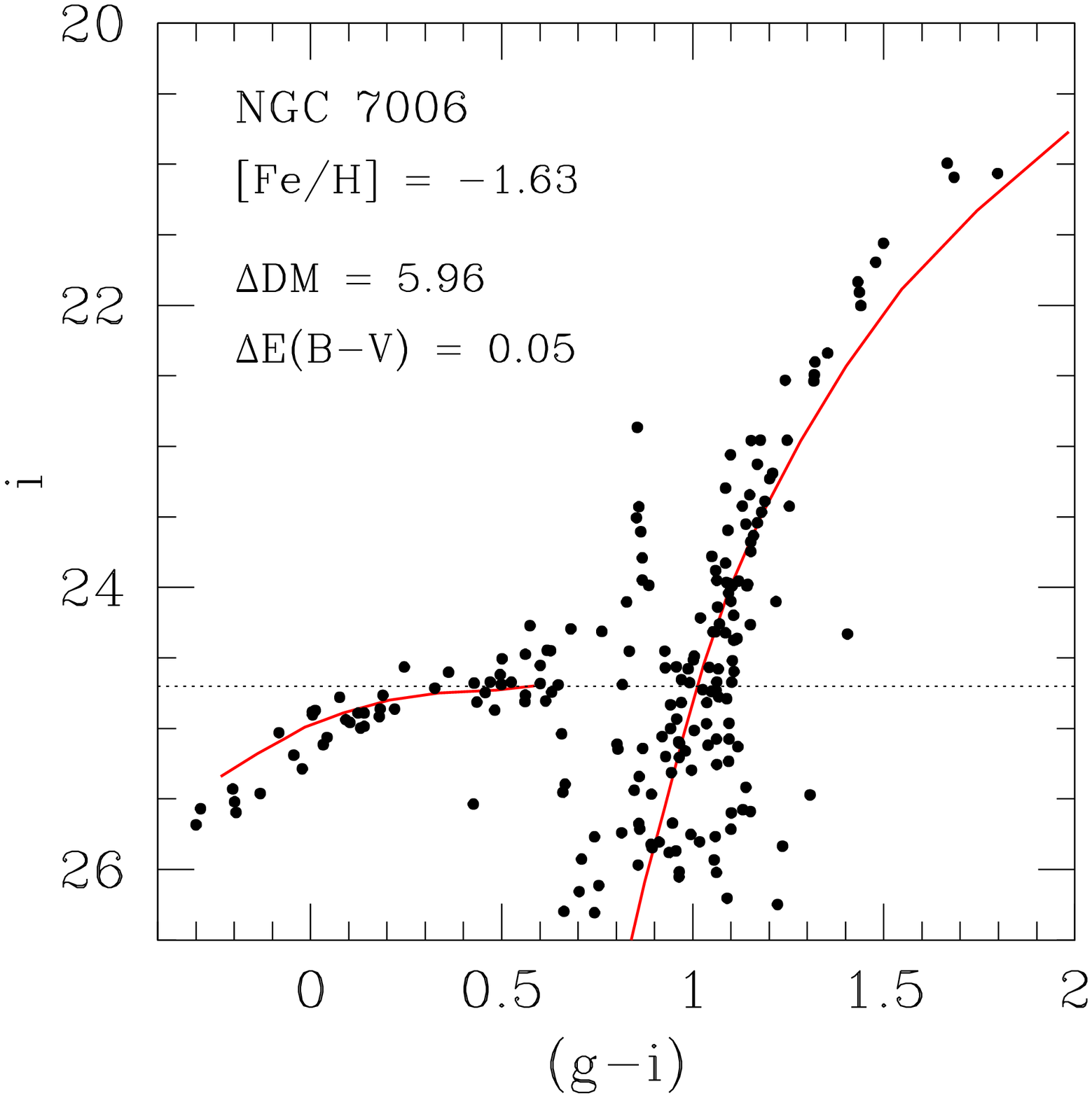}
\hspace{-0.5mm}
\includegraphics[width=0.32\textwidth]{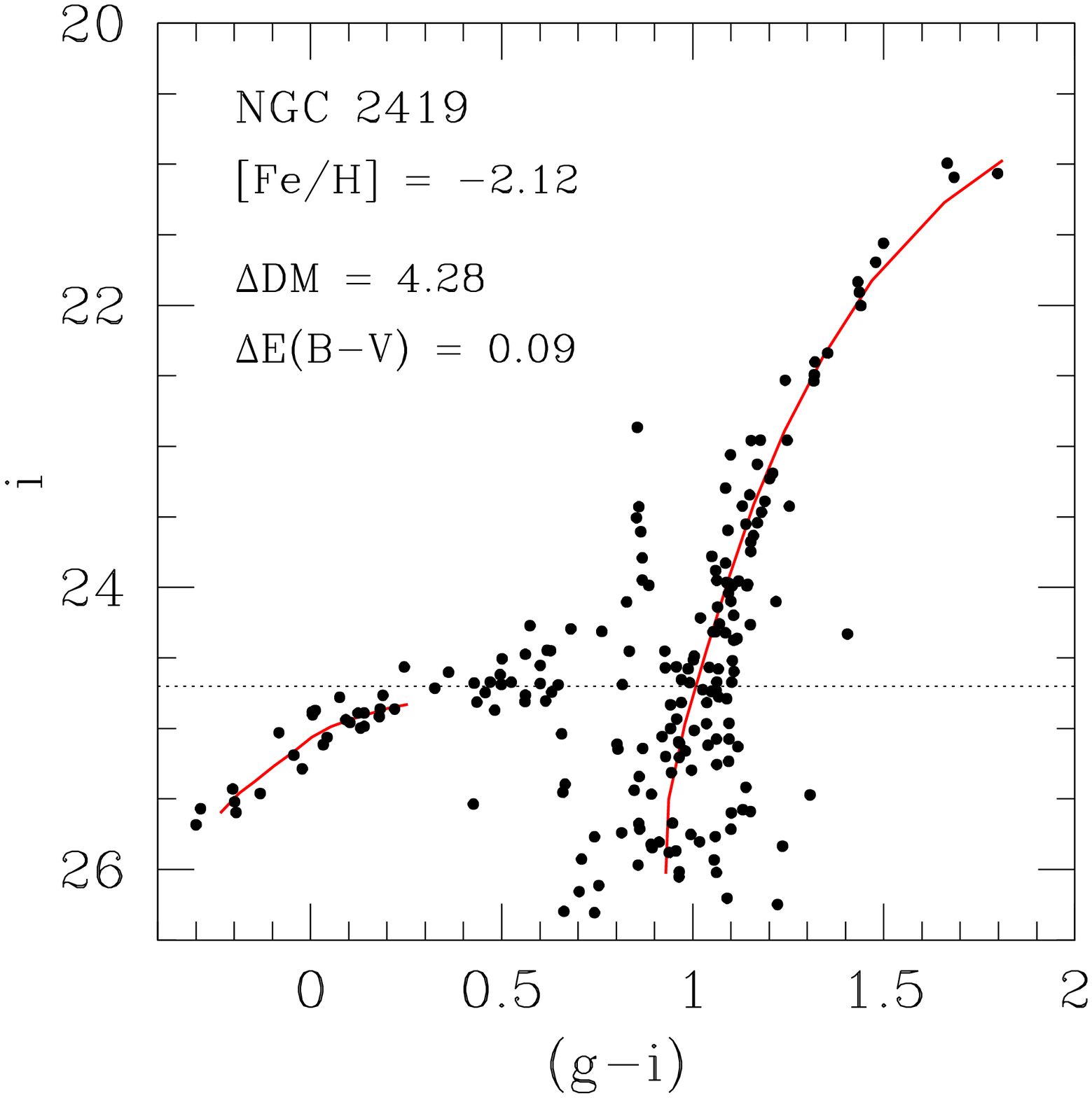}
\hspace{-0.5mm}
\includegraphics[width=0.32\textwidth]{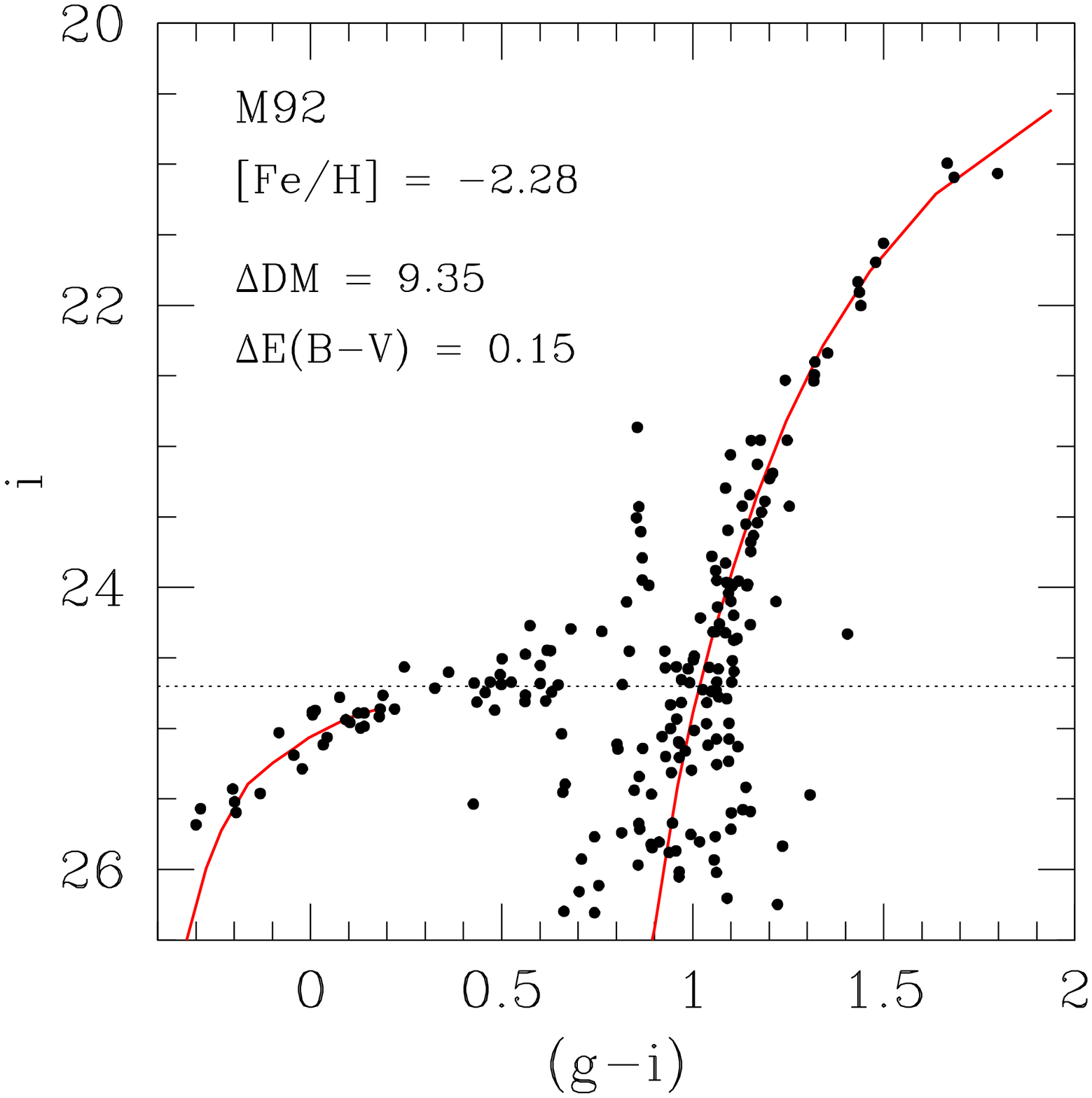}\\
\vspace{1mm}
\includegraphics[width=0.32\textwidth]{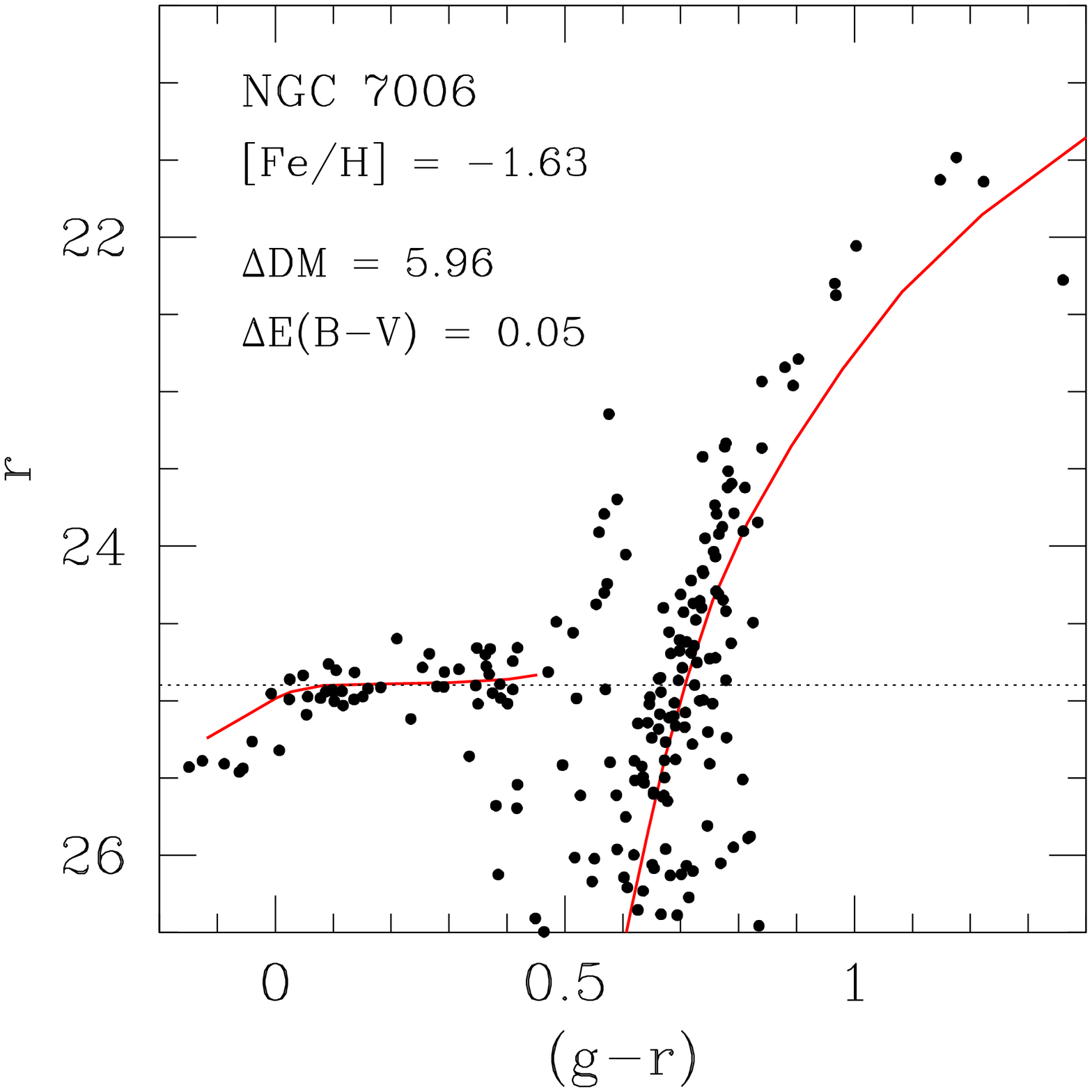}
\hspace{-0.5mm}
\includegraphics[width=0.32\textwidth]{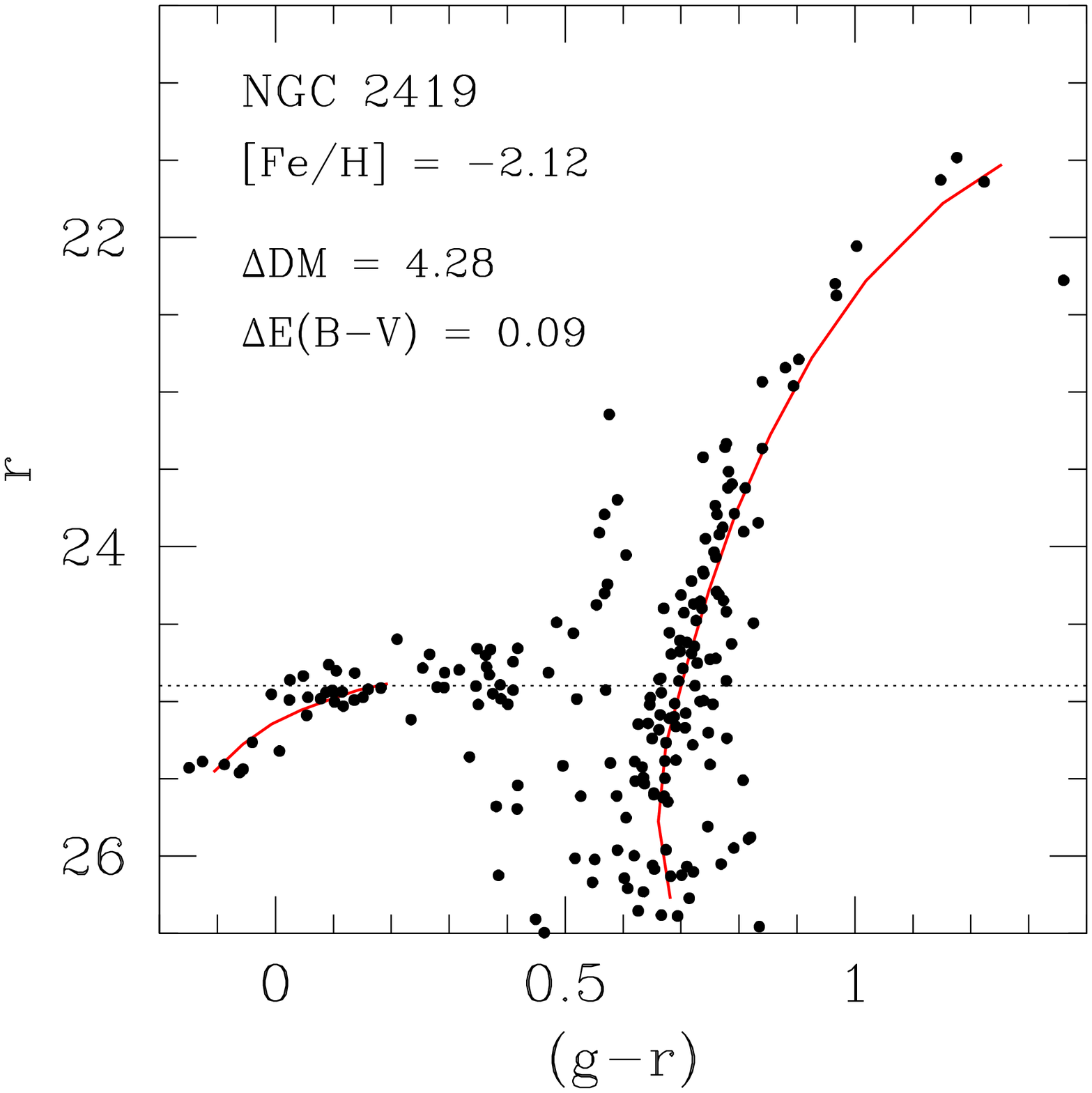}
\hspace{-0.5mm}
\includegraphics[width=0.32\textwidth]{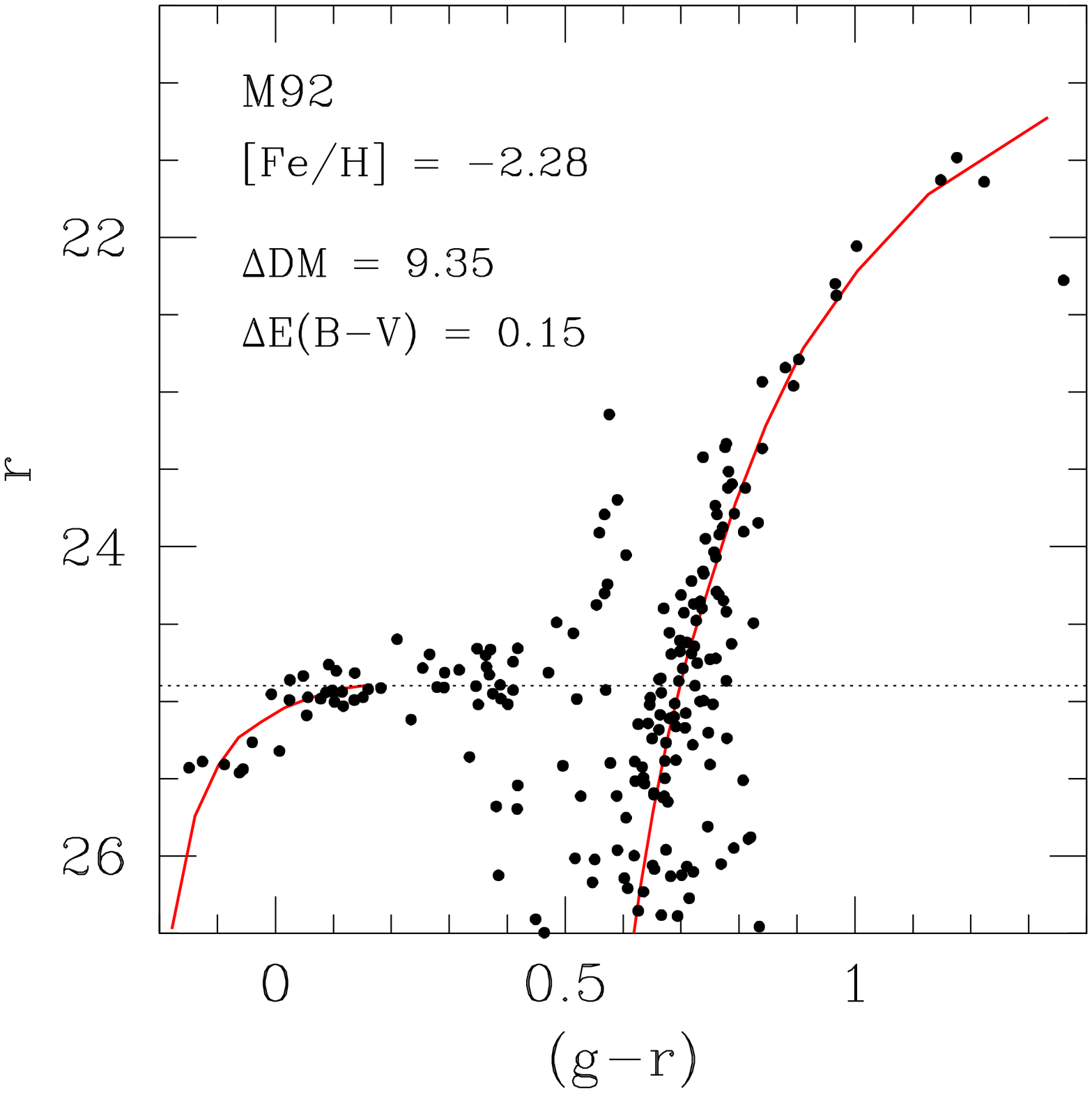}\\
\caption{Fiducial sequences (solid lines) for the Galactic globular clusters NGC 7006 
(left), NGC 2419 (centre), and M92 (right) overlaid on the $(g,i)$ and 
$(g,r)$ CMDs for MGC1. The RGB sequence for NGC 7006 ($[$Fe$/$H$] = -1.63$) is much too red 
to provide a good fit to that of MGC1. In contrast, the sequence for NGC 2419 
($[$Fe$/$H$] = -2.12$) provides a much better fit but diverges slightly to the red 
on the uppermost RGB, while the sequence for M92 ($[$Fe$/$H$] = -2.28$) provides 
an excellent fit along the full extent of the RGB.}
\label{f:metfitggc}
\end{figure*}

The results for NGC 2419 and NGC 7006 are presented in Fig. \ref{f:metfitggc}. The RGB
fiducial for NGC 7006 is clearly much too red, implying that MGC1 is nowhere near as
metal rich as this cluster. In contrast the fiducial for NGC 2419 provides a much better
fit along the majority of the RGB, only diverging marginally to the red towards the top.
This suggests that MGC1 has a metal abundance close to, but still below, that of NGC 
2419 at $[$Fe$/$H$] = -2.12$. 

To explore this further, we used the observations of \citet{clem:08} who provide a fiducial
RGB sequence for the very metal-poor Galactic globular cluster M92 ($[$Fe$/$H$] = -2.28$) 
in the $u\arcmin g\arcmin r\arcmin i\arcmin z\arcmin$ photometric system, which is very similar 
(but not identical) to 
the instrumental SDSS $ugriz$ system used here and by \citet{an:08}. The following relations, 
provided by \citet{tucker:06}, allowed us to transform the \citet{clem:08} M92 fiducial into 
the SDSS $ugriz$ system:
\begin{eqnarray}
g &=& g\arcmin + 0.060\,\left[\left(g\arcmin  - r\arcmin \right) - 0.53\right] \nonumber \\
r &=& r\arcmin + 0.035\,\left[\left(r\arcmin  - i\arcmin \right) - 0.21\right] \label{e:ugriz} \\
i &=& i\arcmin + 0.041\,\left[\left(r\arcmin  - i\arcmin \right) - 0.21\right] \,\,. \nonumber
\end{eqnarray}
\citet{an:08} demonstrate that these transformations are very accurate (better than $2$ per 
cent) for M92 and other Galactic globular clusters. 
Unfortunately, \citet{clem:08} do not provide HB fiducials for their clusters, or a public
archive of their photometry. Therefore, we had to use the HB fiducial for M92 that we
derived from the \citet{an:08} photometry -- this is evidently not ideal, but the demonstrated
accuracy of the transformations listed in Eq. \ref{e:ugriz} means that any systematic
uncertainties introduced should be, at most, a few hundredths of a magnitude. 

The results we obtained by aligning these M92 fiducials are also presented
in Fig. \ref{f:metfitggc}. On both the $(g,i)$ and $(g,r)$ CMDs the agreement is very close
along the full length of the RGB for identical distance modulus and foreground reddening.
It thus seems likely that MGC1 has a very low metal abundance, with $[$Fe$/$H$] \approx -2.3$.
It is worth noting that this value is more metal poor than the initial estimate by
\citet{martin:06}, who obtained $[$Fe$/$H$] \sim -1.3$, and also the value of 
$[$Fe$/$H$] \sim -1.37 \pm 0.15$ measured from high-resolution spectroscopy by 
\citet{alvesbrito:09}. 

In Section \ref{ss:photcal} we noted the lack of colour terms in our photometric
calibration. It is important to assess what impact this may have on the shape of
the RGB and hence on our derived metal abundance. On the $(g,i)$ CMD the part of the 
cluster RGB above the HB spans the colour range $1.0 \la (g-i) \la 1.7$. According 
to Eq. \ref{e:colourterms} the expected systematic offsets due to the omission of
colour terms span the range $0.0 \la \Delta(g-i) \la 0.01$ over this colour interval. 
Similarly, the expected systematic offsets on the $(g,r)$ CMD span the range 
$0.01 \la \Delta(g-r) \la 0.02$. In both cases the result of adding colour
terms is to stretch the RGB to the red, making it appear more metal-rich than in 
the present CMDs. However, the expected magnitude of this effect is too 
small to significantly alter our derived metal abundance.

The horizontal and vertical shifts required to align the M92 fiducials with our CMDs
allow us to estimate the foreground reddening and distance modulus for MGC1. Both
\citet{harris:96} and \citet{schlegel:98} agree that $E(B-V) = 0.02$ for M92, resulting in
an extinction-corrected distance modulus of $\mu = 14.58$ for this cluster \citep{harris:96}. 
The horizontal and vertical offsets we derived, as noted in Fig. \ref{f:metfitggc}, then 
imply that the foreground reddening towards MGC1 is $E(B-V) = 0.17$, and the distance modulus 
for this cluster is $\mu = 23.93$.

There are three additional Galactic globular clusters in the sample of \citet{an:08} that
have metal abundances comparable to that of M92: M15, NGC 5053, and NGC 5466. These
all have truncated RGB sequences and are hence not useful for obtaining a precise 
metallicity estimate for MGC1; however their observed fiducials are none the less
suitable for determining additional estimates of the foreground reddening and distance
modulus of MGC1, on the basis that they are all sufficiently similar in metal abundance
for our alignment procedure to provide physically meaningful results.

\begin{table*}
\caption{Results from fitting metal-poor Galactic globular cluster fiducials to the CMD of MGC1.}
\begin{tabular}{@{}lccccccccccc}
\hline \hline
Cluster & \hspace{1mm} & $[$Fe$/$H$]$$^{a}$ & \hspace{1mm} & $E(B-V)$$^{b}$ & $\mu_0$$^{b}$ & \hspace{1mm} & $\Delta E(B-V)$ & $\Delta \mu$ & \hspace{1mm} & $E(B-V)_{{\rm MGC1}}$ & $\mu_{{\rm MGC1}}$ \\
\hline
NGC 2419 & & $-2.12$ & & $0.09$ & $19.69$ & & $0.09$ & $4.28$ & & $0.18$ & $23.97$ \\
NGC 5466 & & $-2.22$ & & $0.01$ & $15.97$ & & $0.14$ & $7.97$ & & $0.15$ & $23.94$ \\
M15      & & $-2.26$ & & $0.10$ & $15.06$ & & $0.09$ & $8.80$ & & $0.19$ & $23.86$ \\
M92      & & $-2.28$ & & $0.02$ & $14.58$ & & $0.15$ & $9.35$ & & $0.17$ & $23.93$ \\
NGC 5053 & & $-2.29$ & & $0.03$ & $16.10$ & & $0.16$ & $7.75$ & & $0.19$ & $23.85$ \\
\hline
\label{t:ggcfid}
\end{tabular}
\medskip
\vspace{-5mm}
\\
\flushleft
$^a$ Metal abundance as listed by \citet{harris:96} in the 2003 online update. \\
$^b$ Foreground reddening is calculated as the average of the values from \citet{harris:96}
and \citet{schlegel:98}; extinction-corrected distance modulus is calculated using this
value and the apparent visual distance modulus listed by \citet{harris:96}.
\end{table*}

Table \ref{t:ggcfid} summarizes the values of distance modulus and foreground
extinction derived from fitting these cluster fiducials to our CMDs; we have also 
included NGC 2419 for comparison. Random uncertainties in individual values of $\mu$
and $E(B-V)$ are related to the errors associated with aligning the fiducials correctly. 
Experimentation shows that 
fiducials shifted by $\pm 0.02$ mag in colour and in the range $\pm 0.05-0.1$ mag in brightness 
result in noticeably poorer fits. The relationships between $E(B-V)$, and $E(g-i)$ and $E(g-r)$ 
mean that the random uncertainties in $E(B-V)$ are then better than $0.02$ mag. The contribution 
to the uncertainty in $\mu$ from the vertical alignment precision is therefore dominant over 
the contribution from the uncertainty in the calculated extinction, and we hence assume the 
random uncertainty in distance modulus to be $\pm 0.1$ mag. We note that there is also a 
systematic contribution due to the uncertainty in our estimation of the correct HB levels -- 
Fig. \ref{f:lumfunc} suggests this is of order $\pm 0.05$ mag at most.

Our derived colour excesses and distance moduli for MGC1 both exhibit small 
dispersions across the sample of comparison clusters. The rms values are $E(B-V) = 0.18 \pm 0.02$ 
and $\mu = 23.91 \pm 0.06$. We discuss these
results and their implications in more detail in Section \ref{ss:distred}, below.

\begin{figure*}
\centering
\includegraphics[width=0.32\textwidth]{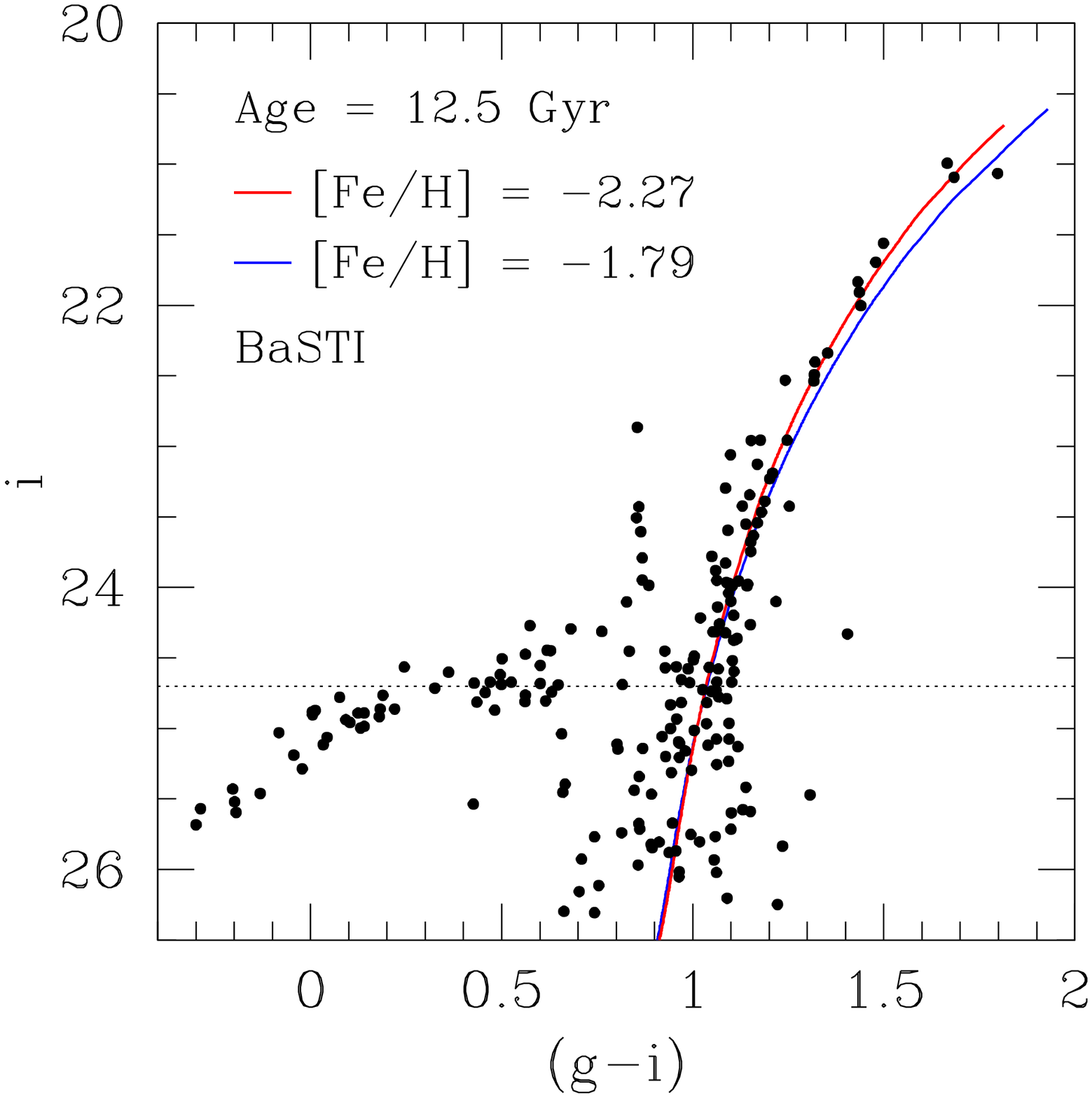}
\hspace{-0.5mm}
\includegraphics[width=0.32\textwidth]{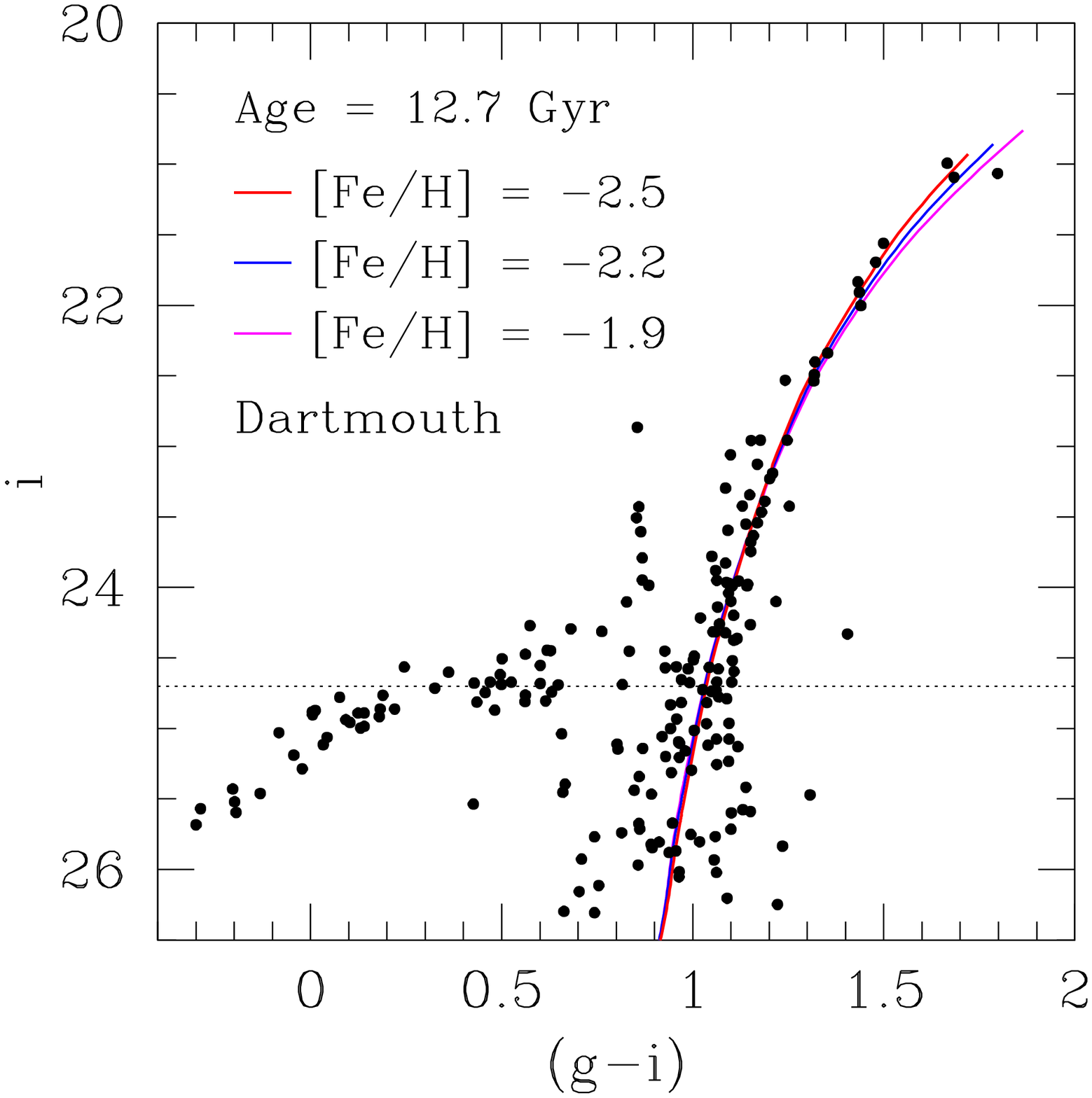}\\
\vspace{1mm}
\includegraphics[width=0.32\textwidth]{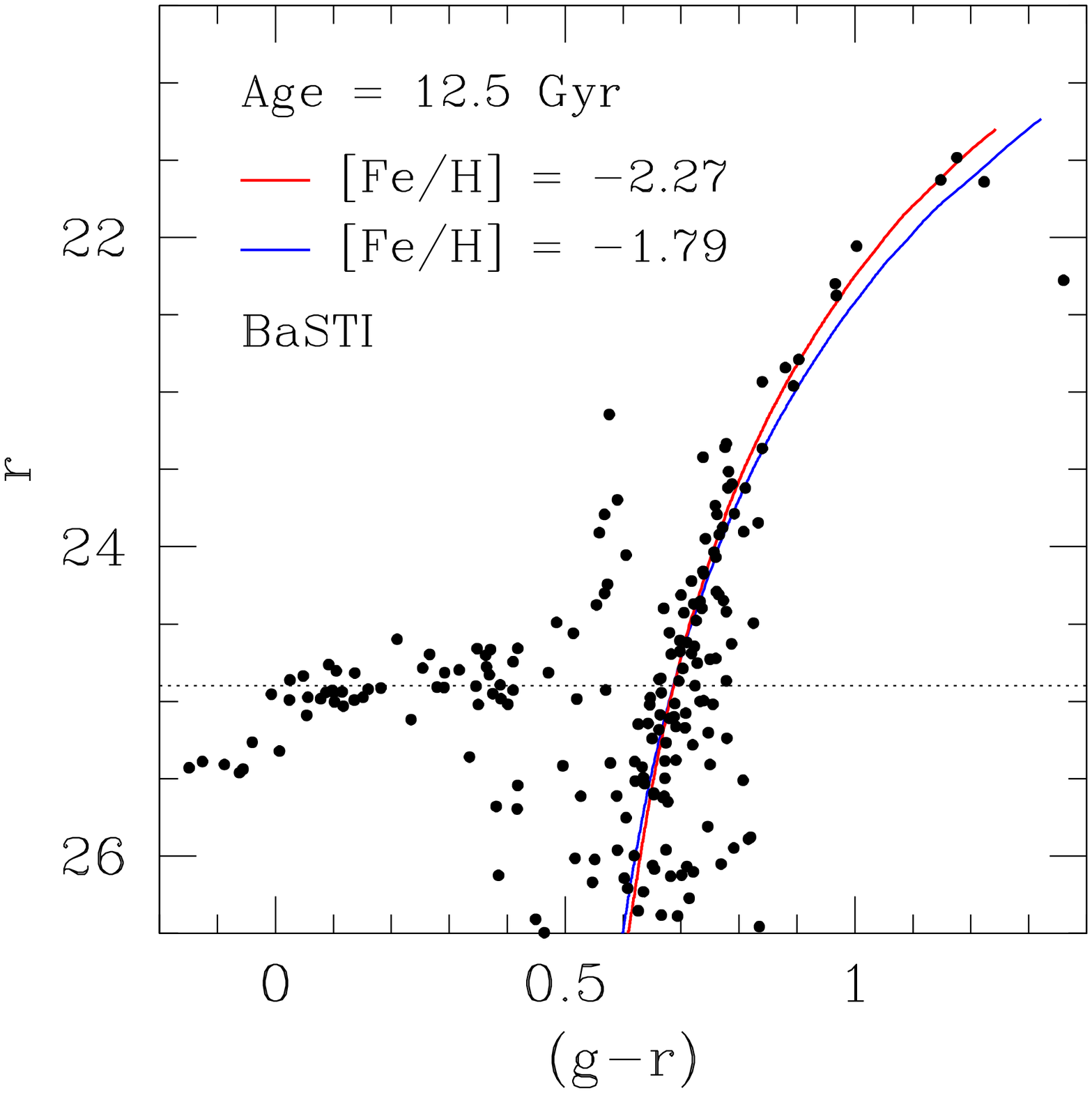}
\hspace{-0.5mm}
\includegraphics[width=0.32\textwidth]{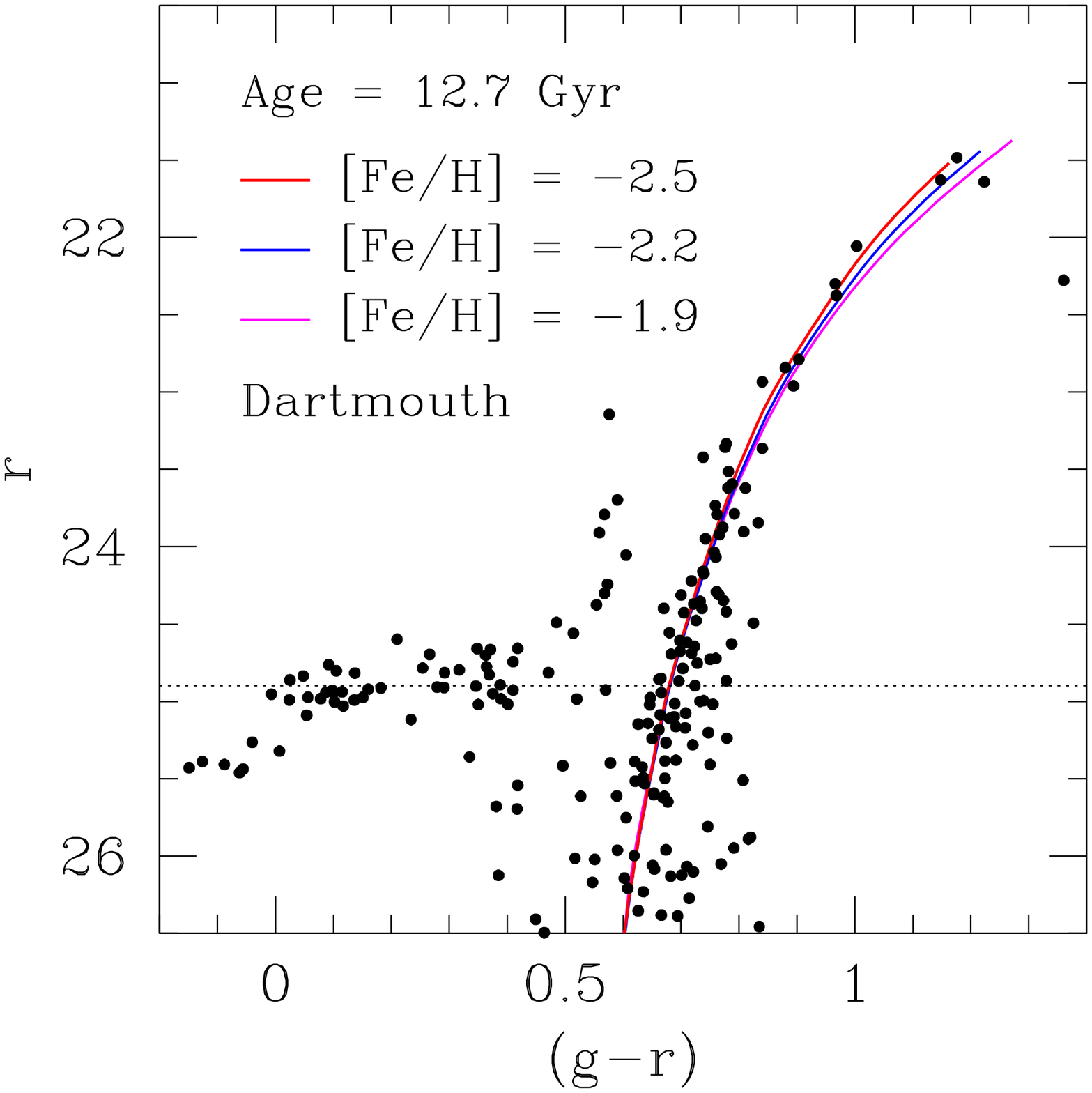}\\
\caption{Red-giant branches from scaled-solar isochrones calculated by the BaSTI and 
Dartmouth groups overlaid on our cluster $(g,i)$ and $(g,r)$ CMDs.
In each panel the isochrone with the best-fitting metallicity is plotted with a thick red line.
The black dotted lines mark the adopted cluster HB level, from Fig. \ref{f:lumfunc}.}
\label{f:metfit}
\end{figure*}

\subsection{Comparison with theoretical isochrones}
To check the consistency of our results, including their robustness to variations in both
the assumed age and $\alpha$-element enhancement of MGC1 (neither of which are tightly
constrained by observation), we employed theoretical isochrones calculated by two different 
groups. We downloaded models of both scaled-solar and $\alpha$-enhanced composition -- 
Galactic globular clusters typically exhibit enhancement in $\alpha$-elements compared 
to the solar abundance, of roughly $[\alpha/$Fe$] \sim +0.3$ \citep[see e.g.,][]{carney:96}.
We initially selected models as close as possible to $\sim 12.5$ Gyr old (taken to be 
representative of the age of the oldest metal-poor Galactic globular clusters), but 
subsequently tested how our results varied with changes in age of $\pm 2$ Gyr either side 
of this value. In more detail, we used the following:
\begin{itemize}
\item{BaSTI models \citep{pietrinferni:04,pietrinferni:06}, for which we downloaded 
isochrones of age $12.5$ Gyr, scaled-solar composition, and metal abundances of 
$[$Fe$/$H$] = -1.79$ and $-2.27$; along with isochrones again of age $12.5$ Gyr,
$\alpha$-enhanced composition with $[\alpha/$Fe$] = +0.4$, and iron abundances of
$[$Fe$/$H$] = -2.14$ and $-2.62$.}\vspace{2mm}
\item{Dartmouth models \citep{dotter:07,dotter:08}, for which we downloaded isochrones of 
age $12.7$ Gyr, metal abundances sampling the range $-2.5 \leq [$Fe$/$H$] \leq -1.6$ at 
$0.3$ dex intervals, canonical Helium abundance  ($Y\approx 0.245$), and both scaled-solar 
and $\alpha$-enhanced composition -- the latter with $[\alpha/$Fe$] = +0.4$. Unlike the
BaSTI isochrones which include post-RGB evolution, the Dartmouth group 
provides separate synthetic HB models.}
\end{itemize}
Although models by the Padova group \citep{girardi:00,girardi:04,marigo:08} are also
commonly employed, we elected not to use these in the present work for two reasons.
First, Padova models with $\alpha$-enhanced composition are not presently available for 
the suspected age and metallicity of MGC1; and second, it is known that the scaled-solar
Padova models in the SDSS $ugriz$ system for ancient metal-poor systems generally possess 
redder RGB loci than should be expected. A particular example of this may be seen in 
\citet{girardi:04} where Padova isochrones in the SDSS photometric system with 
$[$Fe$/$H$] = -1.7$ provide a good fit to the RGB of the Galactic globular cluster NGC 2419, 
which is $\approx 0.4$ dex more metal-poor than this.

We overlaid the isochrones using the same methodology as that described in Section 
\ref{ss:ggc} -- shifting them vertically on the CMD so that their HB levels matched 
that observed for MGC1, and horizontally so that their RGB colours at the level of 
the HB matched the observed value. In this case however, since we were dealing with
theoretical isochrones (which represent the output of stellar evolution models transformed
onto the observational plane), we were forced to relax the additional constraint that the 
implied distance modulus and foreground extinction be identical for both the $(g,i)$ and 
$(g,r)$ CMDs.

\begin{figure*}
\centering
\includegraphics[width=0.32\textwidth]{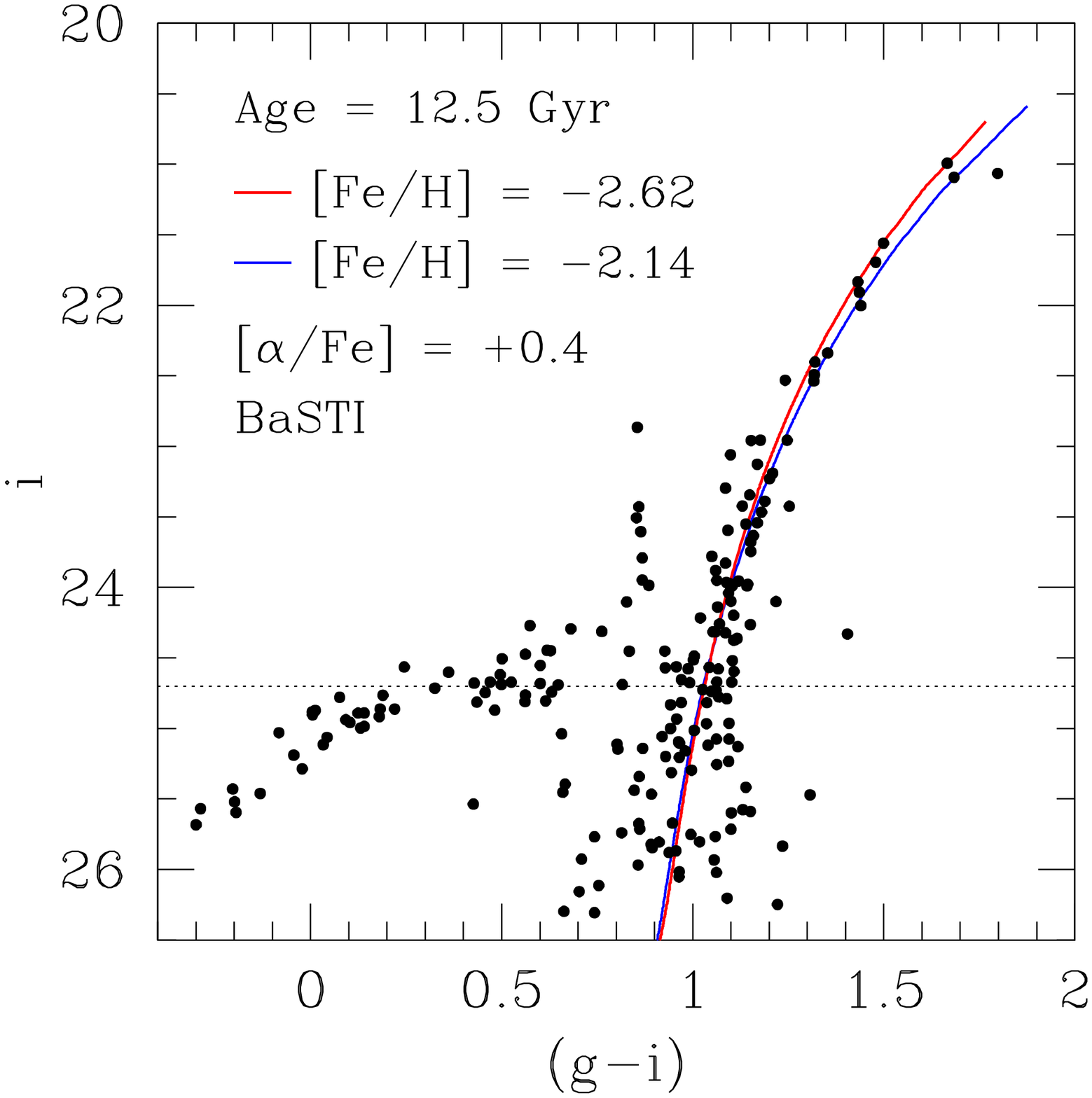}
\hspace{-0.5mm}
\includegraphics[width=0.32\textwidth]{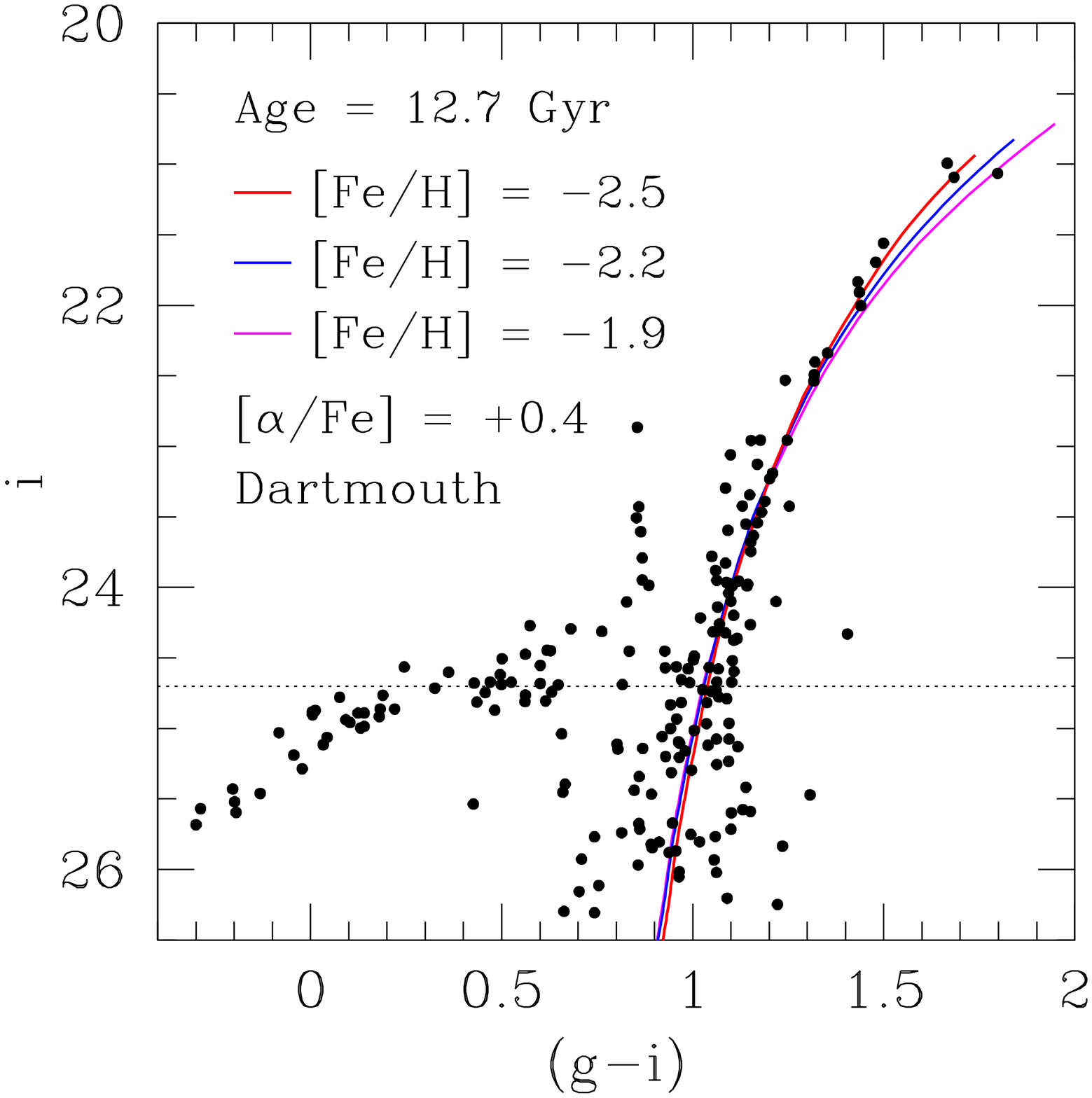}\\
\vspace{1mm}
\includegraphics[width=0.32\textwidth]{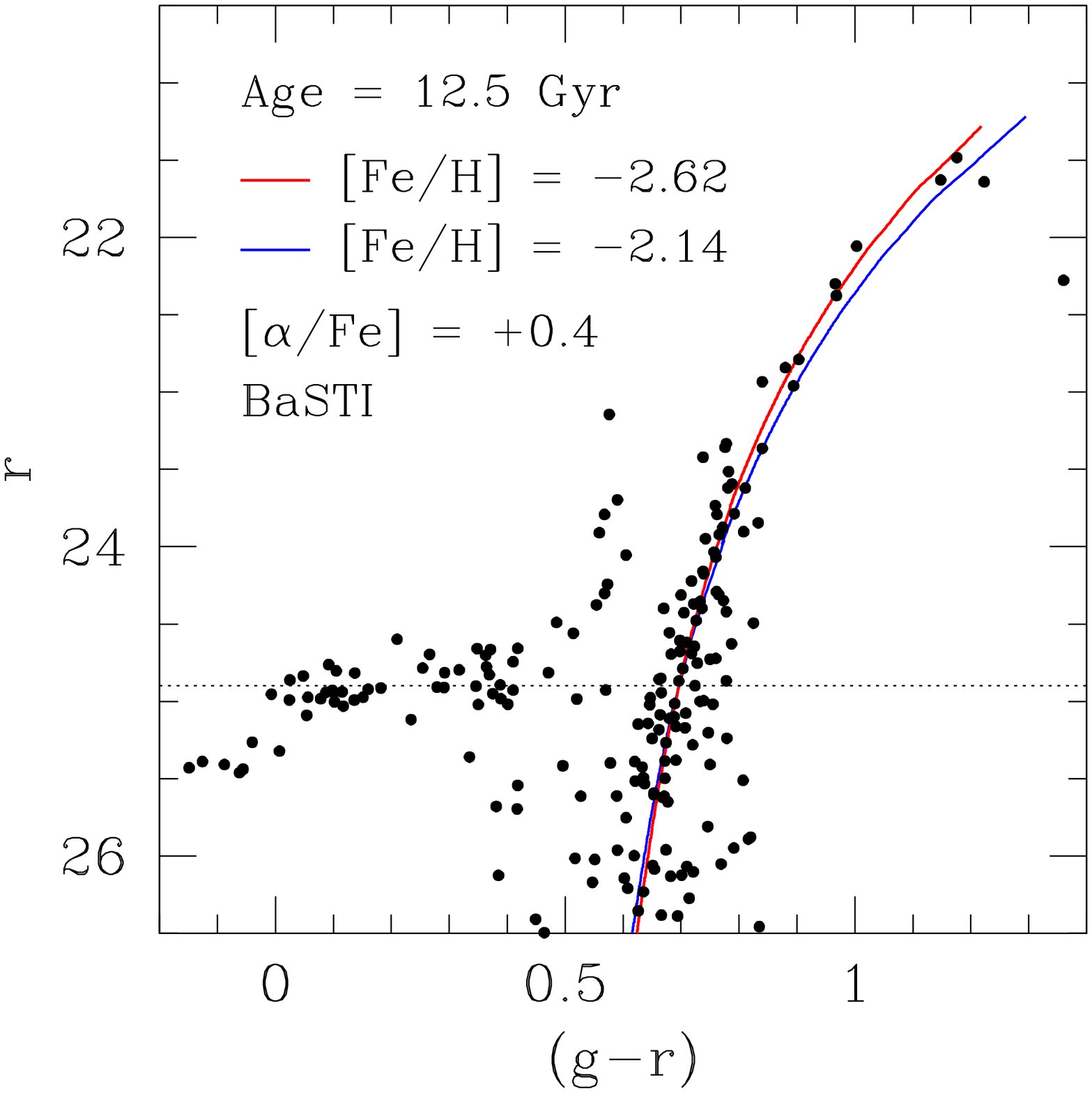}
\hspace{-0.5mm}
\includegraphics[width=0.32\textwidth]{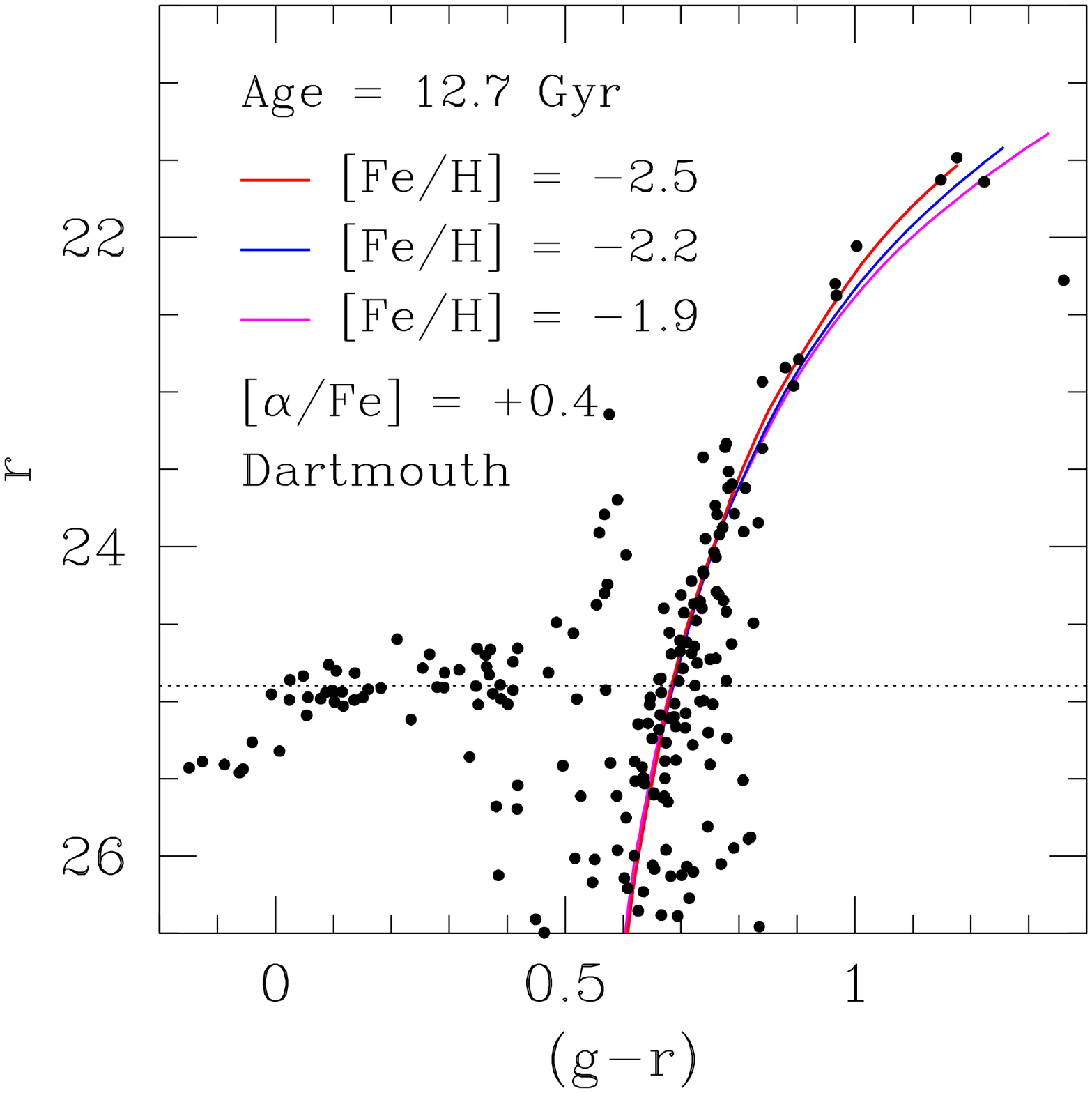}
\caption{Same as Fig. \ref{f:metfit} but for $\alpha$-enhanced isochrones with 
$[\alpha/$Fe$] = +0.4$ from the BaSTI and Dartmouth groups.}
\label{f:metfitae}
\end{figure*}

Fig. \ref{f:metfit} shows aligned scaled-solar 
isochrones with a variety of metallicities on our $(g,i)$ and $(g,r)$ CMDs. 
The results are in excellent agreement with the metallicity estimate of $[$Fe$/$H$] \approx -2.3$
that we obtained by aligning Galactic globular cluster fiducials on the CMD. 
Here, the BaSTI isochrone with $[$Fe$/$H$] = -2.27$ fits the cluster RGB most closely, 
whereas the next available isochrone with $[$Fe$/$H$] = -1.79$ is clearly too metal rich. 
Similarly, the Dartmouth isochrones with $[$Fe$/$H$] = -2.5$ and $[$Fe$/$H$] = -2.2$ both provide 
reasonably good matches to the observed RGB, while the isochrone with $[$Fe$/$H$] = -1.9$ 
is clearly too metal rich. From these models we therefore infer $[$Fe$/$H$] \approx -2.35$. 
To confirm that such measurements are not sensitive to the selected age we repeated the 
alignment procedure using sets of $10$ and $15$ Gyr old isochrones, obtaining consistent 
results to within $\approx 0.15$ dex.

Fig. \ref{f:metfitae} shows aligned $\alpha$-enhanced isochrones on our $(g,i)$ and 
$(g,r)$ CMDs. The degree of $\alpha$-enhancement is marginally higher than the 
$[\alpha/$Fe$] \sim +0.3$ typically observed for Galactic globular clusters and thus likely 
represents the maximum effect $\alpha$-enhancement in MGC1 would have on the metallicity 
estimates we obtained from scaled-solar models. The two $\alpha$-enhanced BaSTI isochrones 
correspond to total metallicities of $[$M$/$H$] = -2.27$ and $-1.79$ \citep{pietrinferni:06}, 
and therefore have lower iron abundances than the scaled-solar models in Fig. 
\ref{f:metfit}. The three $\alpha$-enhanced Dartmouth models, on the other hand, have 
the same iron abundances as the previous scaled-solar models and therefore higher 
total metallicities. Adopting the relation of \citet{salaris:93}:
\begin{equation}
[{\rm M}/{\rm H}] = [{\rm Fe}/{\rm H}] + \log(0.638f_{\alpha} + 0.362)\,\,\,,
\end{equation}
where $f_{\alpha} = 10^{[\alpha/{\rm Fe}]}$ is the $\alpha$-element content, these 
isochrones correspond to total metallicities of $[$M$/$H$] \approx -2.2$, $-1.9$ and $-1.6$, 
respectively. From Fig. \ref{f:metfitae} it is clear that for both the BaSTI and
Dartmouth models it is the most metal poor isochrone that provides the best fit to the 
cluster RGB, implying that if MGC1 has an $\alpha$-enhanced rather than solar-like 
composition it is still a very metal poor object with $[$M$/$H$] \approx -2.2$ or $-2.3$. 
Once again, this is in excellent agreement with the metallicity estimate that we obtained 
by aligning Galactic globular cluster fiducials.

\begin{figure*}
\centering
\includegraphics[width=0.32\textwidth]{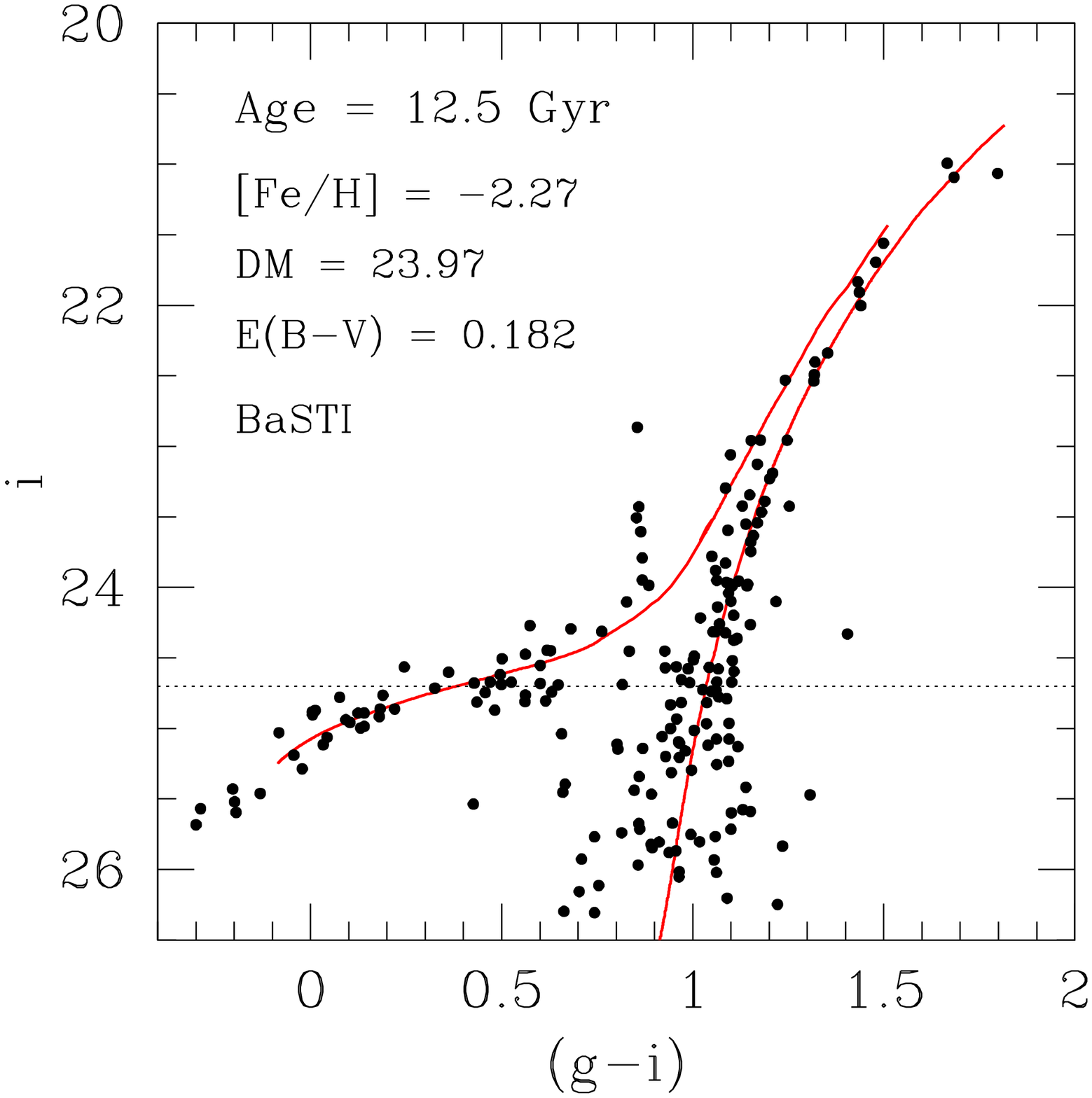}
\hspace{-0.5mm}
\includegraphics[width=0.32\textwidth]{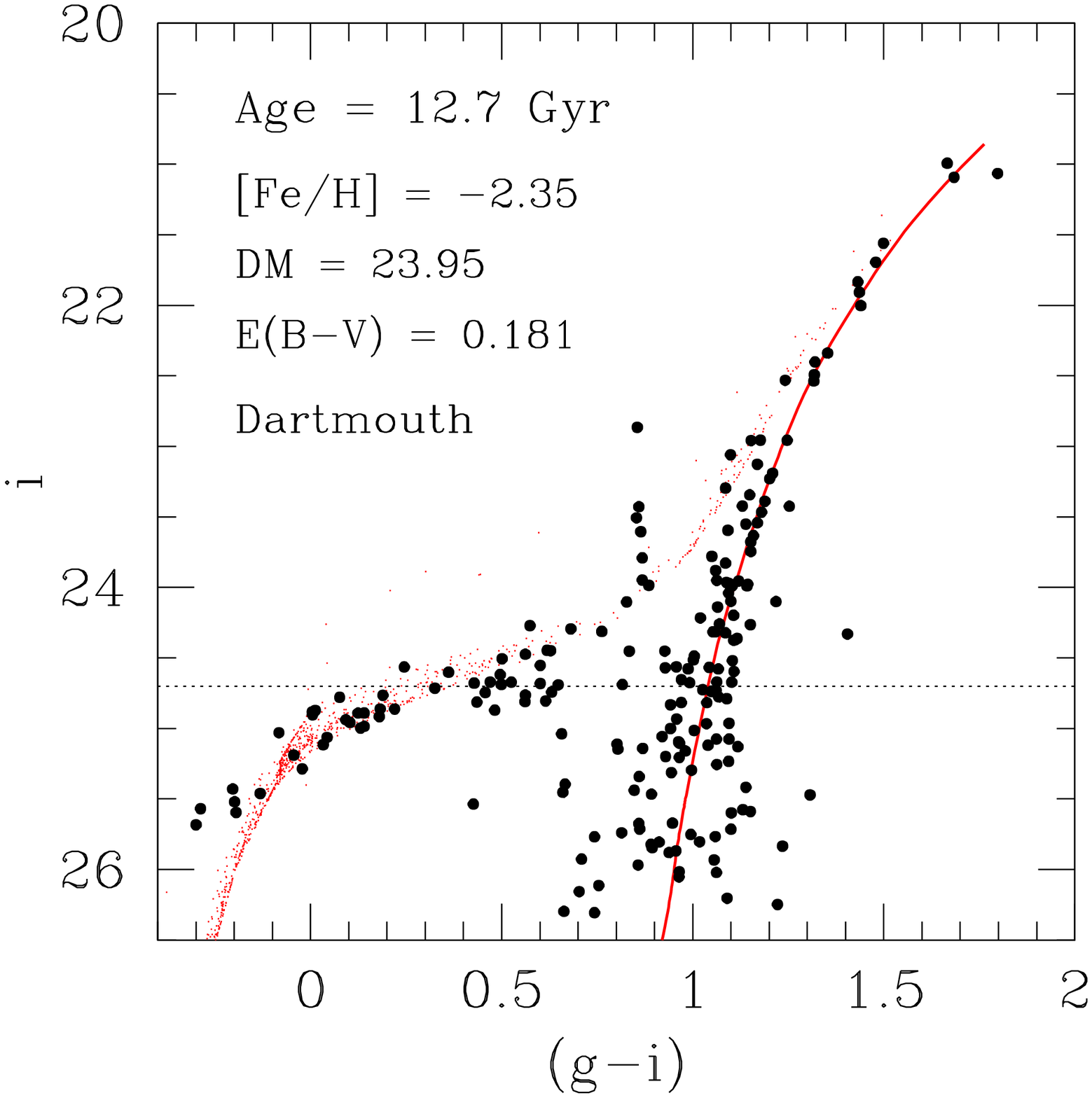}\\
\vspace{1mm}
\includegraphics[width=0.32\textwidth]{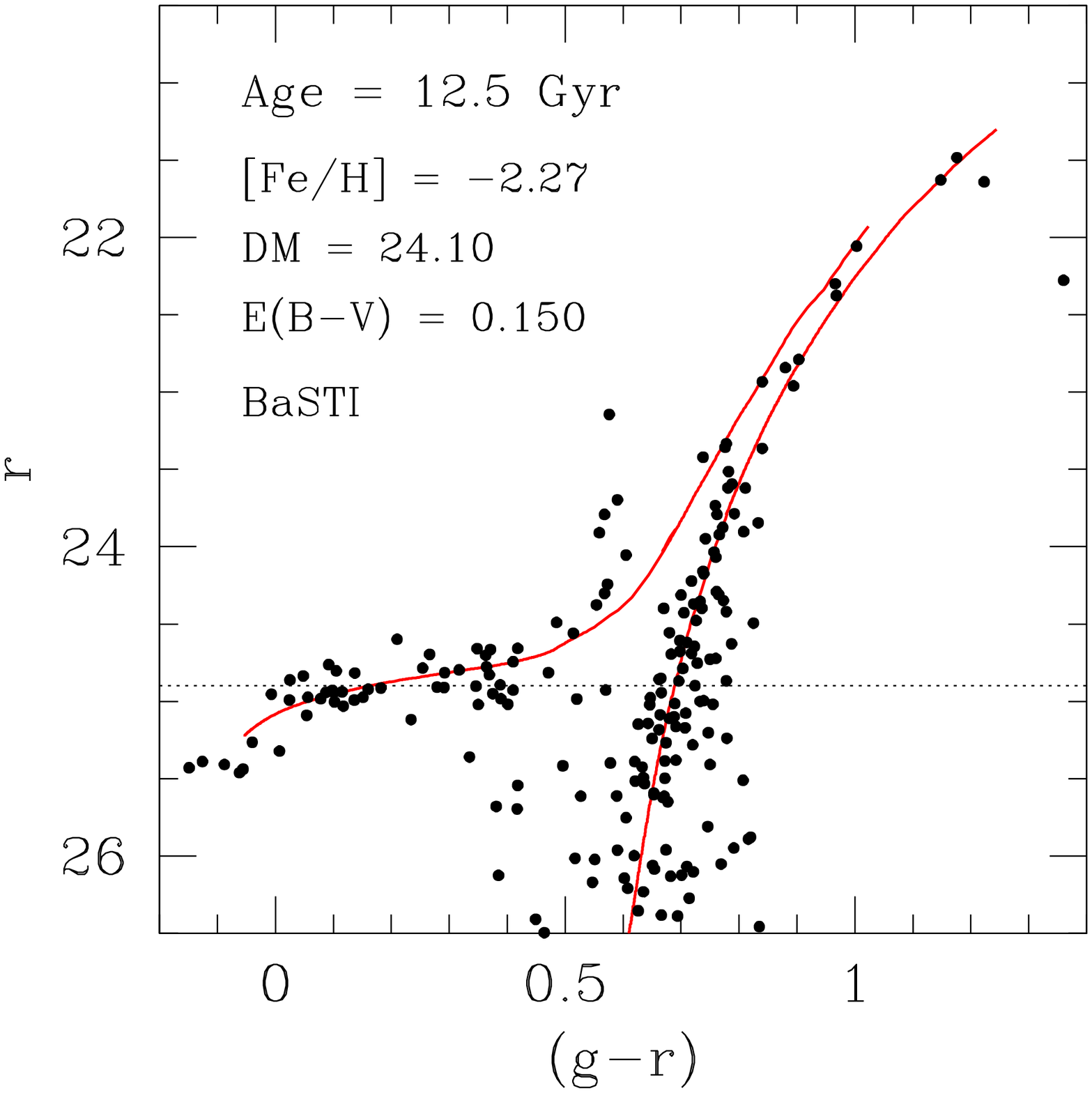}
\hspace{-0.5mm}
\includegraphics[width=0.32\textwidth]{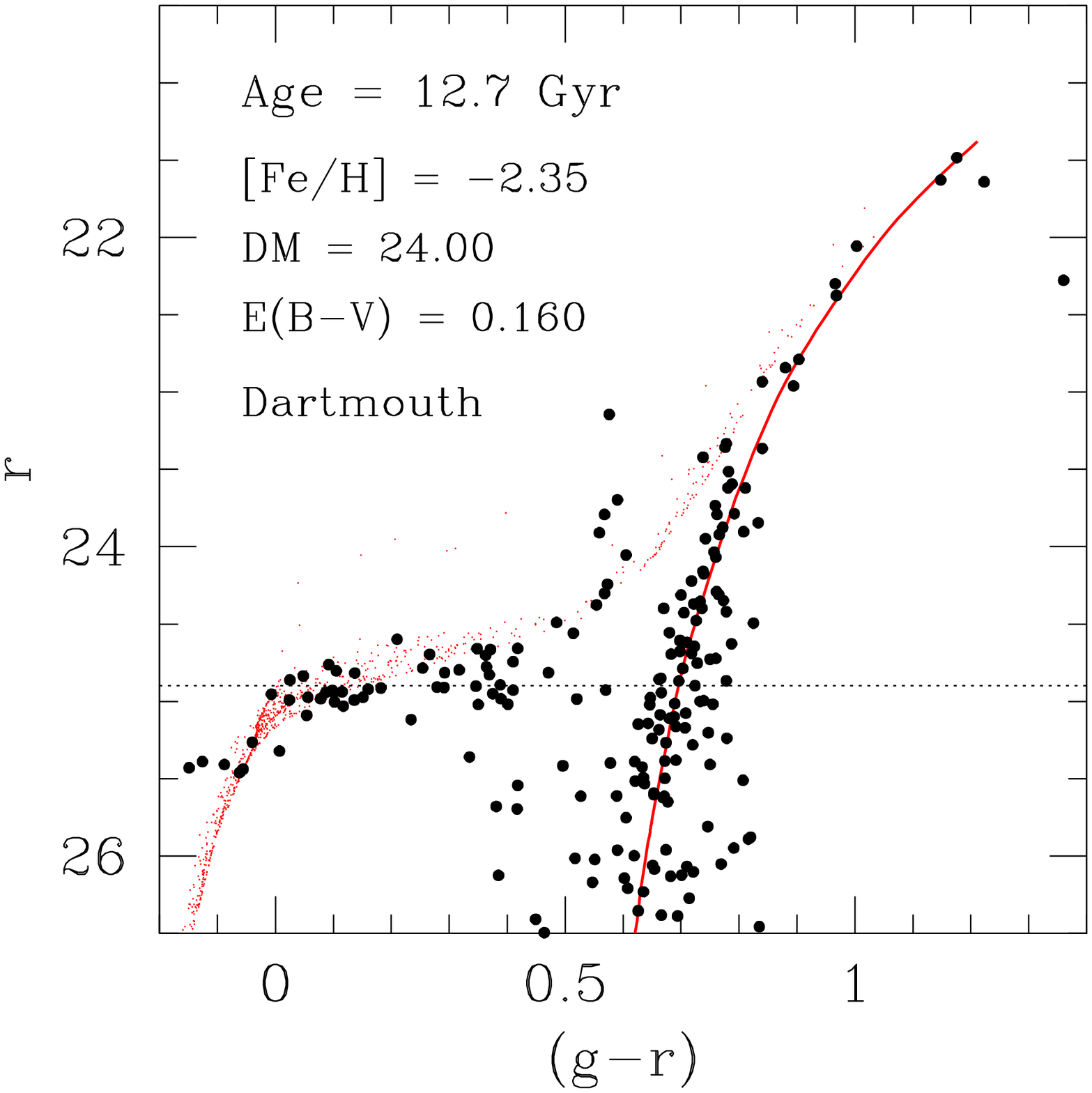}
\caption{CMDs showing the complete best-fit scaled-solar isochrones from 
the BaSTI and Dartmouth groups. The values of $\mu$ and $E(B-V)$ required to align a 
given isochrone are listed in each panel and summarized in Table \ref{t:results}.}
\label{f:isofit}
\end{figure*}

For each set of isochrones we again used the offsets required to align the best-fitting 
model with the CMD to determine estimates of the MGC1 distance modulus and foreground extinction.
Our results are presented in Table \ref{t:results}, while Fig. \ref{f:isofit} shows
examples of the aligned best-fit scaled-solar isochrones from both group. As previously, 
random uncertainties in the derived distance moduli and colour
excesses are related to the errors associated with aligning the isochrones correctly, 
and we assume these to be better than $0.02$ mag in $E(B-V)$ and $\pm 0.1$ mag in $\mu$.

\begin{table*}
\centering
\caption{Results from fitting theoretical isochrones to the CMD of MGC1.}
\begin{tabular}{@{}llcccccccc}
\hline \hline
Isochrone Set & \hspace{2mm} & Filters & \hspace{2mm} & $[$Fe$/$H$]$ & $[\alpha/$Fe$]$ & $[$M$/$H$]$ & \hspace{2mm} & $\mu$ & $E(B-V)$ \\
\hline
BaSTI & & $(g\,,\,i)$ & & $-2.27$ & $0.0$ & $-2.27$ & & $23.97$ & $0.18$ \\
BaSTI & & $(g\,,\,i)$ & & $-2.62$ & $+0.4$ & $-2.27$ & & $23.93$ & $0.19$ \\
BaSTI & & $(g\,,\,r)$ & & $-2.27$ & $0.0$ & $-2.27$ & & $24.10$ & $0.15$ \\
BaSTI & & $(g\,,\,r)$ & & $-2.62$ & $+0.4$ & $-2.27$ & & $24.03$ & $0.17$ \\
\hline
Dartmouth & & $(g\,,\,i)$ & & $-2.35$ & $0.0$ & $-2.35$ & & $23.95$ & $0.18$ \\
Dartmouth & & $(g\,,\,i)$ & & $-2.50$ & $+0.4$ & $-2.21$ & & $24.01$ & $0.18$ \\
Dartmouth & & $(g\,,\,r)$ & & $-2.35$ & $0.0$ & $-2.35$ & & $24.00$ & $0.16$ \\
Dartmouth & & $(g\,,\,r)$ & & $-2.50$ & $+0.4$ & $-2.21$ & & $24.08$ & $0.15$ \\
\hline
\label{t:results}
\end{tabular}
\end{table*}

The derived values of distance modulus and foreground reddening listed in Table
\ref{t:results} show good consistency irrespective of the adopted isochrone set 
or $\alpha$-element abundance. More specifically, the calculated distance moduli 
span the range $23.93-24.10$ with an rms value of $\mu = 24.00 \pm 0.06$, while 
the $E(B-V)$ measurements span the range $0.15-0.19$ with an rms value of $0.17 \pm 0.02$. 
These results are entirely consistent with those we derived by fitting Galactic 
globular cluster fiducials to our CMDs.

\subsection{MGC1 distance and foreground extinction}
\label{ss:distred}
Overall, our best estimate for the distance modulus of MGC1 is $23.91 \pm 0.06$ from fitting
Galactic globular cluster fiducials to our CMD, and $24.00 \pm 0.06$ from fitting theoretical 
isochrones. These correspond to line-of-sight distances of $605$ kpc and $630$ kpc respectively, 
indicating that MGC1 lies substantially nearer to us than do the central regions of M31, which 
have a line-of-sight distance of $\approx 780$ kpc 
\citep[$\mu_{{\rm M31}} = 24.47 \pm 0.07$;][]{mcconnachie:05}. This result is consistent with 
the previous distance estimate for MGC1 by \citet{martin:06}.

\begin{figure*}
\begin{minipage}{175mm}
\begin{center}
\includegraphics[width=95mm]{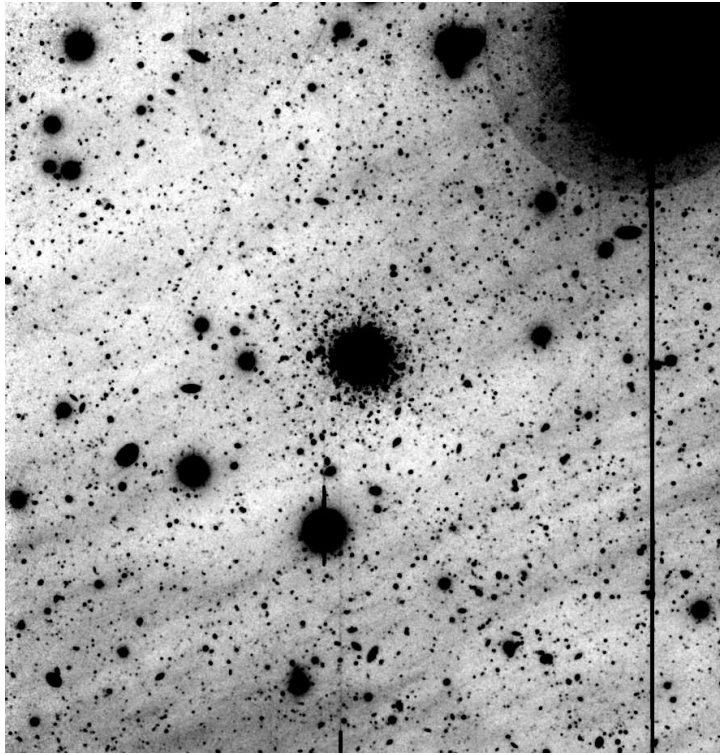}
\caption{Subaru/Suprime-Cam $g$-band cut-out of MGC1, spanning $6\arcmin \times 6\arcmin$,
showing the spatially variable nebulosity found in this part of the sky.
The image has been smoothed slightly with a Gaussian kernel of FWHM\ $\sim 2$ pixels
($\approx 0.4\arcsec$) and then had an appropriate linear scaling applied to show the local
excess nebulosity clearly. North is towards the top of the image and east is to the left.}
\label{f:nebulosity}
\end{center}
\end{minipage}
\end{figure*}

The distance we have measured for MGC1 allows us to estimate the three-dimensional 
(deprojected) distance between this cluster and the centre of M31. In doing so, we adopt 
$\mu_{{\rm M31}} = 24.47 \pm 0.07$ along with a corresponding projected distance between 
the cluster and M31 of $R_{{\rm p}} = 117$ kpc. The derived deprojected galactocentric 
radius of MGC1 is then $205$ kpc for a distance modulus of $23.91$ and $R_{{\rm gc}} = 185$ kpc
for $\mu = 24.00$. Recalling that the $1\sigma$ uncertainties on our measured distance moduli 
are $\pm 0.06$ mag, the uncertainties in these deprojected radii are $\pm 20$ kpc.

Meanwhile, our estimate of the MGC1 colour excess is $E(B-V) = 0.18 \pm 0.02$ from fitting
Galactic globular cluster fiducials to our CMD, and $E(B-V) = 0.17 \pm 0.02$ from fitting
theoretical isochrones. These values are a factor of two larger than the prediction from 
the reddening maps of \citet{schlegel:98} for this region of the sky: $E(B-V) = 0.086$.
Because this discrepancy exists for all of the different Galactic globular cluster fiducials
we considered and for all of the different isochrone sets we used, and because it occurs 
consistently on both the $(g,r)$ and $(g,i)$ CMDs, it is unlikely to be a result of our 
data reduction or analysis. It also cannot be due to the omission of colour terms in 
determining our photometric calibrations, as we discuss below.

A possible explanation for our large inferred $E(B-V)$ value may be seen from
examination of wide-field imaging of the region surrounding the cluster. 
Fig. \ref{f:nebulosity} shows a $6\arcmin \times 6\arcmin$ $g$-band image centered 
on the cluster, excised from Subaru Suprime-Cam observations targeting the dwarf 
galaxy And XIII (Martin et al. 2009, in prep.) which lies projected nearby to
MGC1. The image shows nebulosity varying across the region, implying that there 
is variable extinction present on scales much smaller than the 
$6\arcmin \times 6\arcmin$ resolution of the \citet{schlegel:98} maps. 
In particular, inspection of Fig. \ref{f:nebulosity}
reveals that the excess nebulosity happens to be large on the cluster, consistent
with our derived colour excess. We note that a similar pattern is also visible on the 
MegaCam survey imaging of this region from which MGC1 was discovered, and also, to some 
extent, on our GMOS imaging. This consistency indicates that the nebulosity visible in 
Fig. \ref{f:nebulosity} is a real feature of this part of the sky, and not simply due to 
a flat-fielding error or to scattered light on the Subaru image.

It is worthwhile considering the effects that the omission of colour terms from our
photometric calibration (see Section \ref{ss:photcal}) might have on our derived distance 
modulus and foreground extinction. These quantities were measured by registering fiducial
sequences to the cluster HB luminosity and RGB colour. On the 
$(g,i)$ CMD, the level region of the HB occurs at $(g-i) \approx 0.5$. Eq. 
\ref{e:colourterms} then suggests that leaving out colour terms in our photometric calibration
results in an $i$-band HB magnitude that is too faint by $\sim 0.01$ mag. In other words,
incorporating this correction would make the distance modulus shorter by $\sim 0.01$ mag.
Similarly, the colour of the RGB at the HB level is $(g-i) \approx 1.0$. Again
using Eq. \ref{e:colourterms}, at this colour the effect of omitting colour terms
from our photometric calibration is negligible, $\Delta (g-i) \approx 0.00$.
For the RGB above the HB level, the effect of omitting colour terms is to produce 
an RGB that is too blue by up to $\sim 0.01$ mag. Thus, including the colour terms
would only act to increase our derived colour excess. Following the same process for
the $(g,r)$ CMD, we obtain very similar results. Overall, it is then clear that
(i) including colour terms in our photometric calibration would not have resulted
in our deriving a larger distance modulus or smaller foreground extinction for
MGC1; and (ii) the expected corrections are, in any case, negligible in size.

\section{Cluster structure and luminosity}
We determined the radial density profile of MGC1 in each of our three photometric
passbands via a combination of surface photometry in its unresolved central regions, 
and star counts at larger radii. We obtained surface photometry by using the 
{\sc phot} task in {\sc iraf} to measure the flux through circular apertures of 
increasing size, and then converting these into measurements of the surface brightness 
in concentric circular annuli. An estimate of the sky background was determined from 
several blank regions near the cluster, and subtracted from each annulus. We desired
to extend the integrated-light photometry as far from the cluster centre as possible 
to maximize overlap with star-count profiles (see below); in practice we were 
limited to a maximum radius $r_{{\rm p}} = 35\arcsec$,
beyond which the flux contribution from objects not belonging to the cluster became
non-negligible (as determined from the CMD). Even within this radius there were
several significant background galaxies evident on the images (see Fig. \ref{f:cluster});
these were masked prior to our measurements. The surface photometry 
was transformed onto the SDSS photometric system via Eq. \ref{e:inst2sdss}.

\begin{figure}
\centering
\includegraphics[width=0.47\textwidth]{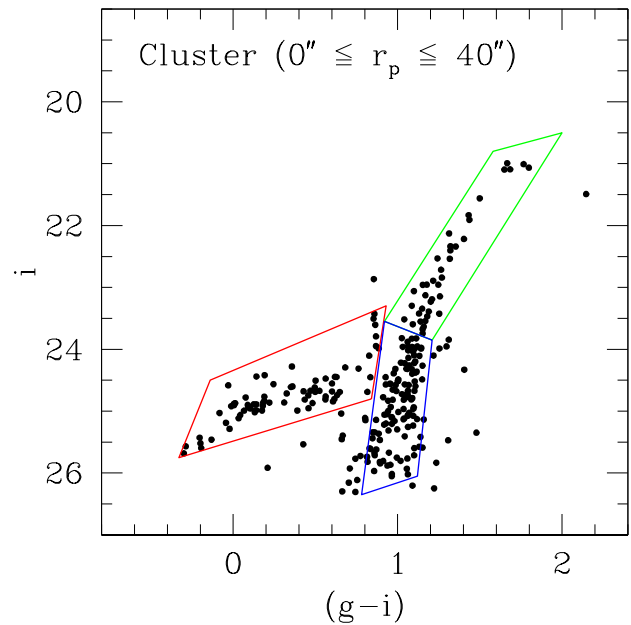}
\caption{All detected stars within $40\arcsec$ of the centre of MGC1 plotted on a
$(g,i)$ CMD. These sequences define the three selection boxes (marked) that
we used to identify probable cluster members at larger radii.}
\label{f:selection}
\end{figure}

We employed star counts to extend the radial profile beyond $35\arcsec$. To minimize
contamination from non-cluster members we only counted stars falling in several 
empirically-determined regions on both the $(g,i)$ and $(g,r)$ CMDs. On each CMD we used 
stellar detections within $40\arcsec$ of the centre of MGC1 to define three boxes 
preferentially inhabited by cluster members -- one region enclosing the HB, and one each 
for the lower and upper RGB. Because we suspected variable foreground 
extinction on scales of a few arcminutes or less (see Section \ref{ss:distred}), we 
ensured these CMD regions were sufficiently broad to accommodate changes in $E(B-V)$ 
of up to $\pm 0.1$ mag away from the cluster centre. Example selection
boxes for the $(g,i)$ CMD can be seen in Fig. \ref{f:selection}.

Our star-count annuli extended from the innermost regions of MGC1 out to the edges 
of our stacked images at $r_{{\rm p}} = 150\arcsec$. To ensure consistency with the
inner surface photometry, in each individual annulus we converted the summed flux
of the counted stars into a surface brightness. Because this process did not include
unresolved diffuse cluster light, there was a clear offset between our star-count
profile and that from the inner surface photometry. To join the two profiles we
considered an overlap region, extending inwards from $35\arcsec$ to a radius where
we judged that incompleteness due to crowding was having a significant effect on
the star-count profile. This was clearly visible as a point near 
$r_{{\rm p}} \approx 13 \arcsec$ where the shapes of the two profiles began to 
diverge -- that from the star counts flattening and falling below that from the 
surface photometry. 

\begin{figure*}
\centering
\includegraphics[width=0.45\textwidth]{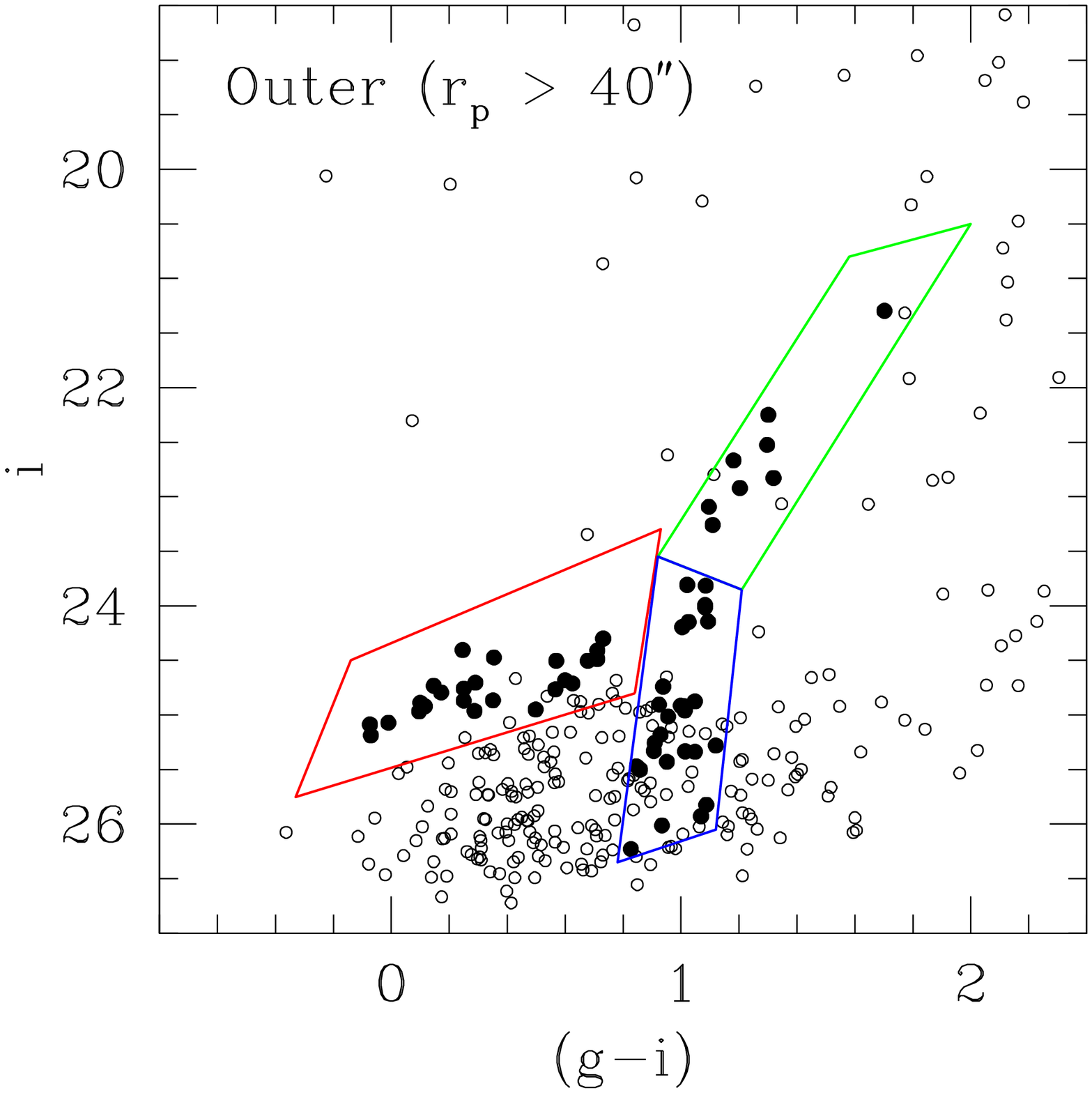}
\hspace{-0.5mm}
\includegraphics[width=0.45\textwidth]{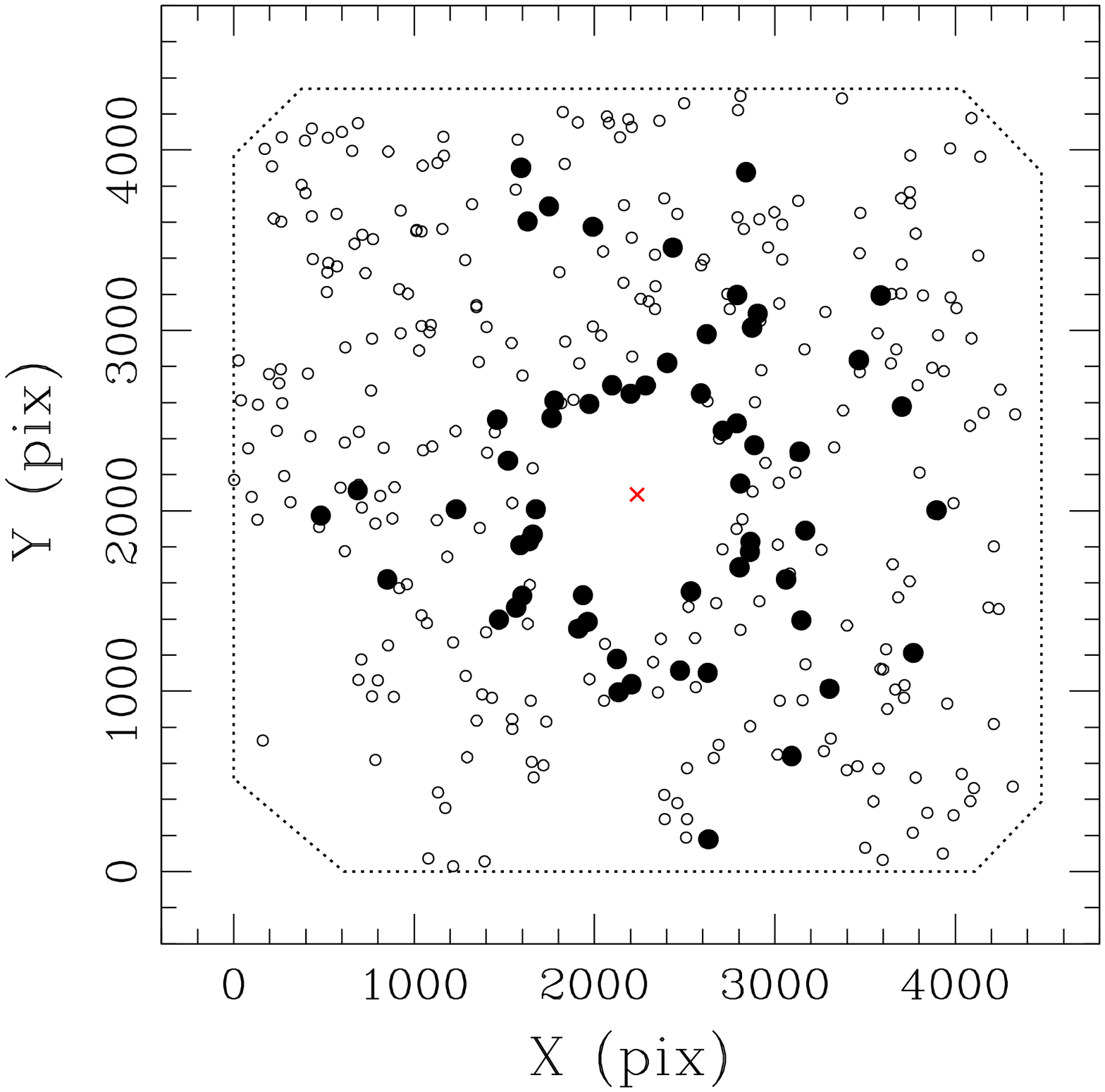}
\caption{All stellar objects in our GMOS field outwith $40\arcsec$ from the centre of
MGC1. {\it Left panel:} Plotted on a $(g,i)$ CMD along with the selection boxes from 
Fig. \ref{f:selection}. {\it Right panel:} Spatial positions on the GMOS field of view. 
In each panel, stars which fell within selection boxes on both the $(g,i)$ and $(g,r)$ 
CMDs are marked with solid black circles; all other stellar objects are marked with
open circles. In the
left panel there are clear overdensities of stars visible at the blue end of the HB and 
halfway up the RGB, indicating the presence of cluster members at large radii. 
The right panel shows that these members extend to the edge of the field of view (marked
with dotted lines) and appear to be fairly evenly distributed about the cluster centre
(marked with a cross).}
\label{f:outer}
\end{figure*}

Although our CMD selection boxes greatly reduced contributions to the star-count
profile from non-cluster objects, there was still undoubtably sufficient contamination
that it was necessary for us to correct for this. We originally planned to use the 
outermost regions of our GMOS imaging to define the background level; however, to our 
surprise the star-count profile was still clearly falling steeply as a function of radius 
to the very edges of the field -- indicative of a significant contribution from cluster 
members even at large radii. Evidence of this may be seen in Fig. \ref{f:outer}, the
left panel of which shows a CMD for all stellar objects at radii beyond $40\arcsec$. 
Objects falling in the selection boxes on both the $(g,i)$ and 
$(g,r)$ CMDs are marked with large solid circles. Although there is certainly 
contamination in the $(g,i)$ CMD selection boxes from non-cluster members, there is 
also a significant number of stars falling very close to where the main cluster sequences 
are expected. In particular a clear overdensity of stars is visible at the blue end of 
the HB, as well as around the join between the lower and upper RGB. Many of these objects 
fall at radii even beyond $80\arcsec$. A plot of their spatial positions (right panel of
Fig. \ref{f:outer}) reveals them to be fairly evenly distributed about the cluster 
-- no striking azimuthal asymmetry is evident.

To obtain an estimate of the background level and trace the outermost regions of the
cluster profile we utilised two supplementary wide-field data sets -- photometry from 
the CFHT/MegaCam survey imaging in which MGC1 was discovered \citep[see][]{martin:06,ibata:07}, 
and photometry from the Subaru Suprime-Cam observations noted in Section \ref{ss:distred} 
(see Martin et al. 2009, in prep.). 
The CFHT observations are in the MegaCam $g$ and $i$ passbands\footnote{Which
are similar, but not identical, to the SDSS $g$ and $i$ filters.} and are roughly 
$1.5-2$ mag shallower than our GMOS observations. The Subaru observations are intermediate
in both depth and image quality between our GMOS observations and the CFHT imaging.
Both the Subaru/Suprime-Cam and CFHT/MegaCam photometric data sets were obtained as 
end-products from versions of the CASU reduction pipeline \citep{irwin:01}. Both cover
regions of sky to radii at least $\sim 2000\arcsec$ from the centre of MGC1.

To utilize these data sets we followed a similar procedure to that for constructing 
our GMOS star-count profile. We first defined appropriate selection boxes on the 
respective $(g,i)$ CMDs by using stars within $40\arcsec$ of the cluster centre, although 
we were careful to (i) exclude the most crowded inner regions in both sets where the photometry 
is degraded, and (ii) not include objects too close to the faint limit in the Subaru 
photometry, which is spatially variable over the full field of view. We then used these 
selection boxes to perform star counts in concentric circular annuli extending to 
$400\arcsec$ from the cluster centre. We determined the background level by considering a 
wide annulus exterior to this, spanning $r_{{\rm p}} = 400-2000\arcsec$. We joined the two
profiles to our composite GMOS profile by considering the overlap in the radial range
$20-150\arcsec$. Note that the MegaCam photometry is not strictly on the
same system as the GMOS and Subaru photometry; however assuming there is no strong
colour gradient in the outer regions of the MGC1 profile, the difference is, to a good
approximation, a small constant offset in magnitude\footnote{{\scriptsize {\it http://www2.cadc-ccda.hia-iha.nrc-cnrc.gc.ca/megapipe/docs/filters.html}}}.

\begin{figure*}
\centering
\includegraphics[width=0.49\textwidth]{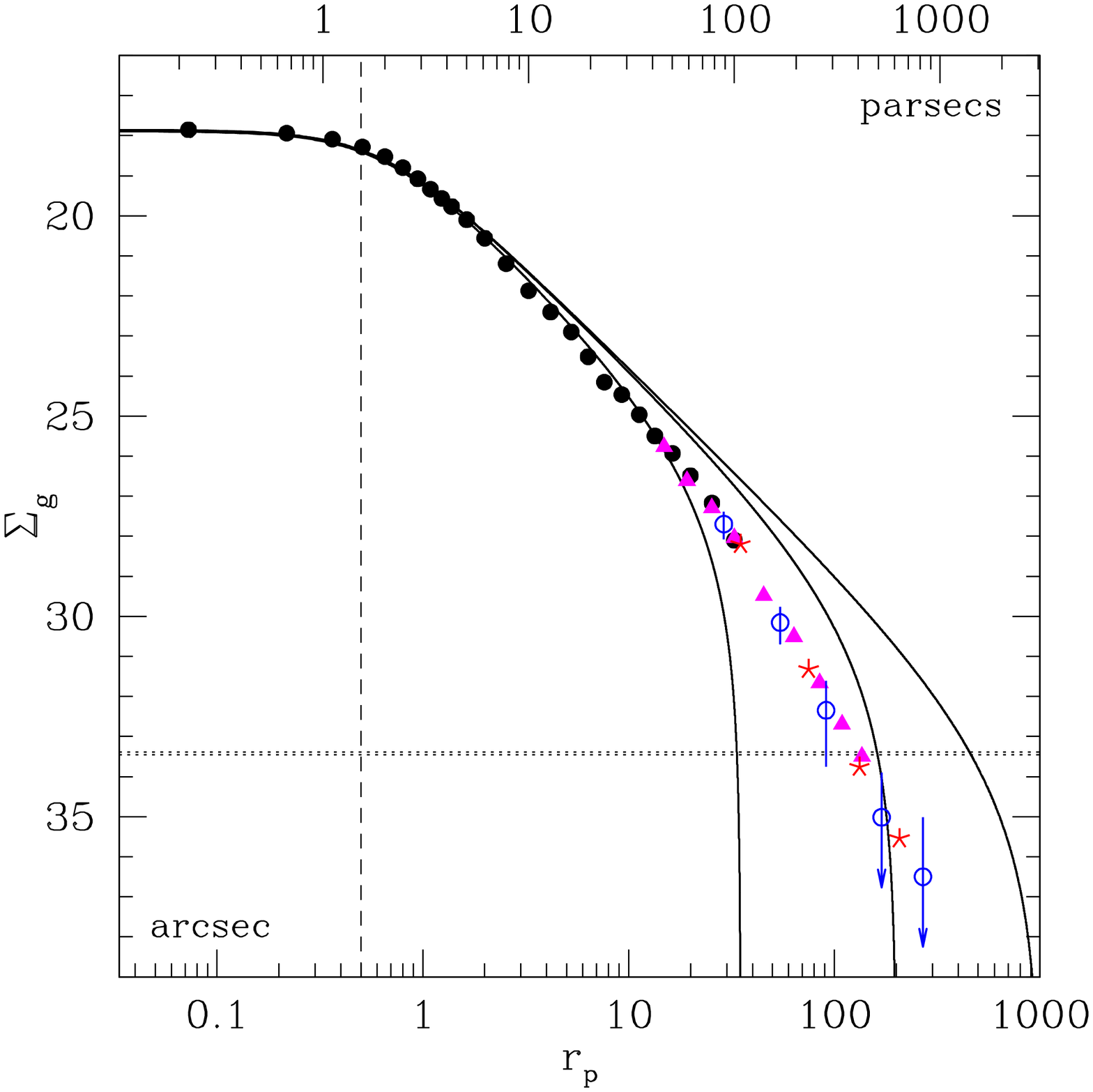}
\hspace{-0.5mm}
\includegraphics[width=0.49\textwidth]{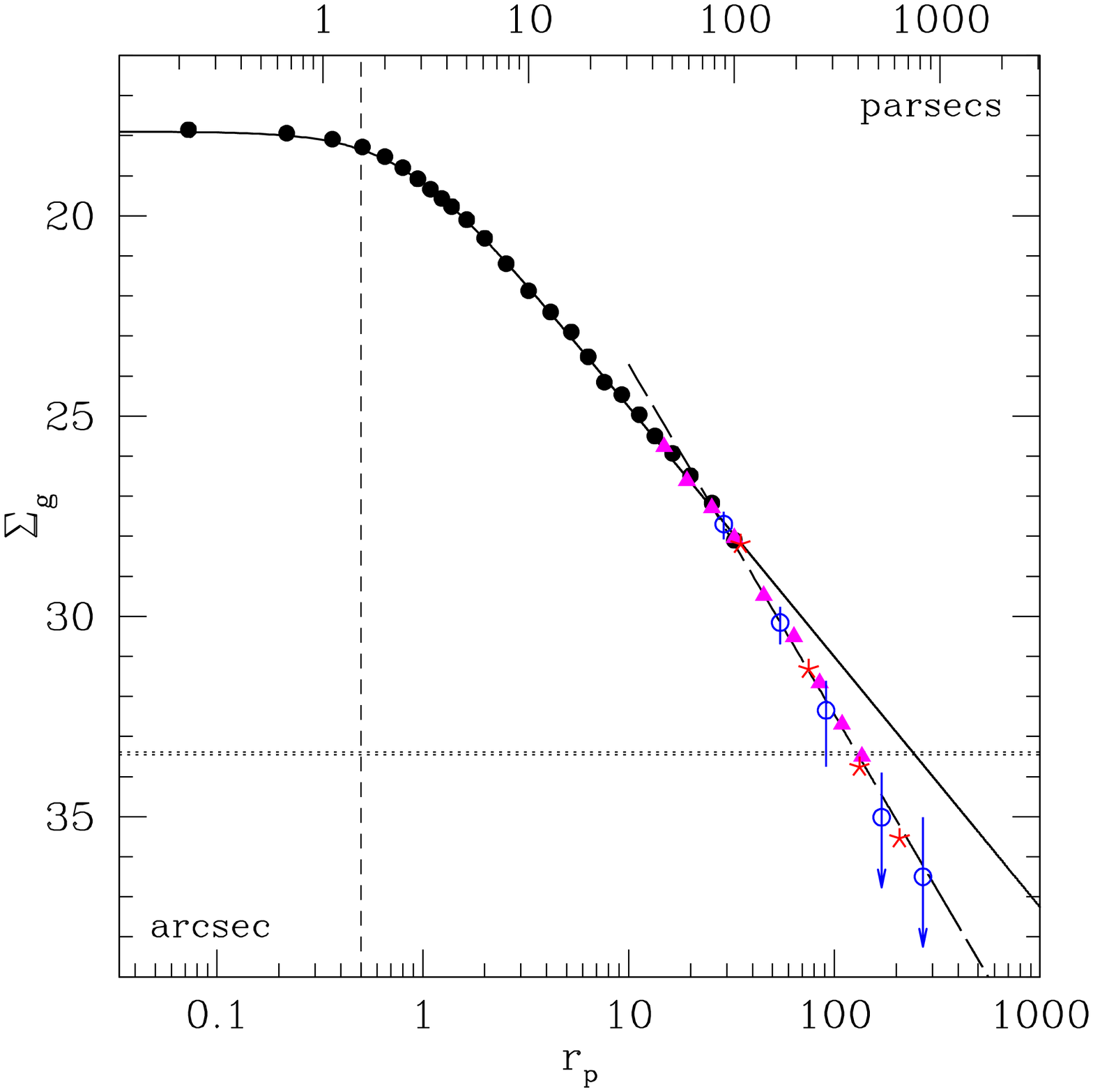}
\caption{Radial background-subtracted SDSS $g$-band surface brightness profile for MGC1. 
The solid black circles indicate integrated-light photometry in the unresolved central regions
from our GMOS imaging, while the solid magenta triangles denote star-count measurements
at larger radii from the same observations. At larger radii still, star counts from our 
CFHT/MegaCam survey imaging are indicated by red stars, while star counts from 
Subaru/Suprime-Cam are marked by open blue circles. Uncertainties are plotted on the
Subaru points; at given radius these are typical for all the star-count measurements.
Background levels derived from the MegaCam and Suprime-Cam observations are marked with
horizontal dotted lines; these agree to better than $0.1$ mag. The vertical dashed line
indicates the $0.5\arcsec$ FWHM of the $g$-band PSF; at smaller radii atmospheric
seeing strongly affects the profile. The left panel shows
three \citet{king:62} models with $r_c = 0.65\arcsec$ and $r_t = 35\arcsec$, $200\arcsec$ 
and $1000\arcsec$. The right panel shows our best-fitting EFF model, which has $a = 0.8\arcsec$
and power-law fall-off $\gamma = -2.5$. There is a break to a steeper power-law fall-off
with $\gamma = -3.5$ (long-dashed line) around $r_{{\rm p}} \sim 35\arcsec$.}
\label{f:structure}
\end{figure*}

The final $g$-band background-subtracted radial brightness profile for MGC1 may be seen 
in Fig. \ref{f:structure}. The background levels measured from the MegaCam and Subaru
photometry agree to within $0.1$ mag and are marked by the horizontal dotted lines.
We determined uncertainties in the central surface photometry by splitting a given
annulus into up to $8$ segments and considering the variance in flux between them;
the resulting error-bars are smaller than the sizes of the plotted points. Uncertainties 
in the star count profiles were calculated using a combination of Poisson statistics and
the uncertainty in the background subtraction. To avoid congestion
in the Figure we have marked error-bars on the Subaru annuli only, however these 
are indicative of the typical uncertainty at a given radius. The vertical dashed line
marks the FWHM of the GMOS $g$-band PSF and is thus the approximate radius interior to
which atmospheric seeing strongly affects the cluster profile.

The GMOS measurements alone (solid black circles and magenta triangles) show that the
cluster can be reliably traced to a radius $r_{{\rm p}} = 150\arcsec \approx 450$ pc
(assuming $\mu = 23.95$), at which point the surface brightness is comparable to the
background contribution. Points exterior to this are strongly sensitive to the adopted
background level, but, taken at face value, the profile apparently extends to 
$\sim 250-300\arcsec \approx 750-900$ pc without showing a clear tidal cut-off.

We attempted to fit, by $\chi^2$ minimization, two sets of commonly-used models
to the MGC1 profile. 
The first were empirical \citet{king:62} models of the form:
\begin{equation}
\Sigma(r_{{\rm p}}) = \Sigma_0 \left[ \frac{1}{\sqrt{1 + \left( \frac{r_{{\rm p}}}{r_c} \right)^2}} - \frac{1}{\sqrt{1 + \left( \frac{r_t}{r_c} \right)^2}} \right]^2
\end{equation}
where $r_t$ is the tidal cut-off radius, $r_c$ is the core radius, and $\Sigma_0$ is 
a scaling parameter. As long as one is not dealing with a low-concentration cluster
(ie, provided $r_t \gg r_c$), $\Sigma_0$ is approximately equal to the central surface 
brightness, and $r_c$ is roughly the radius where the surface brightness has fallen to
half its central value.

We also fit models of the form described by \citet*[][hereafter EFF]{elson:87}:
\begin{equation}
\Sigma(r_{{\rm p}}) = \Sigma_0 \left( 1 + \frac{r_{{\rm p}}^2}{a^2} \right)^{-\frac{\gamma}{2}}
\end{equation}
which are similar to the \citet{king:62} models in their central regions but have
no outer tidal cut-off, instead approaching a power-law fall-off with exponent $-\gamma$
at large radius. Here, the scale radius $a$ is related to the King core radius $r_c$ by:
\begin{equation}
r_c = a \sqrt{2^{\frac{2}{\gamma}} - 1}
\end{equation}
provided, once again, that $r_t \gg r_c$.

The results of our model fitting are displayed in Fig. \ref{f:structure}. The left panel
shows that while a King profile fits the inner regions of MGC1 well, in the
outer parts these models do not provide a good description of the observed shape of
the profile, even if a wide variety of tidal radii are invoked (the three examples
plotted have $r_c = 0.65\arcsec$ and $r_t = 35\arcsec$, $200\arcsec$ and $1000\arcsec$). 
The EFF family of models does somewhat better, although a single set of parameters cannot 
describe the full radial extent of the observations. The cluster apparently closely follows 
a power-law fall-off in its outer regions, but there is a break around 
$r_{{\rm p}} \sim 35\arcsec$ from a profile with exponent $\gamma \approx -2.5$ to a 
steeper fall-off with $\gamma \approx -3.5$. It is notable that the break occurs
roughly where our composite profile transitions from surface photometry to star counts,
and a stellar luminosity function that steepens with radius could thus potentially be 
responsible. However, the fact that the break can apparently be seen when considering 
only the GMOS star counts alone (magenta triangles) argues against this interpretation.

Our best-fit models allow us to precisely estimate the cluster core radius: the King 
profiles have $r_c = 0.65\arcsec \approx 2.0$ pc, while the EFF model has
$a = 0.80\arcsec$ which corresponds to $r_c = 0.69\arcsec \approx 2.1$ pc 
(assuming $\gamma = -2.5$, which is most appropriate for the inner part of the observed 
profile). Both sets of models converge to a central surface brightness 
$\Sigma_g(0) = 17.85$ mag$/$arcsec$^{2}$. Our measured values for $r_c$ are
close to the FWHM of the PSF for the $g$-band observations, and could therefore
conceivably be smaller than we have measured here. Such a scenario would also result
in a larger central surface brightness than we have determined. Higher resolution 
observations of MGC1 will be required to test this possibility. For now we simply
note that many globular clusters with smaller $r_c$ and higher central surface 
brightness than we have measured for MGC1 are known both in the inner and outer regions 
of M31 \citep*[e.g.,][]{barmby:02,tanvir:09}.

Our observed profile also allows an estimation of the cluster luminosity.
Integrating the measured surface brightness out to the edge of the GMOS field at
$r_{{\rm p}} = 150\arcsec$, and subtracting the Subaru/MegaCam background level
results in a total apparent luminosity $m_g = 15.74$\footnote{It is unnecessary
to integrate to larger radii, where the background-subtracted profile is quite 
uncertain, as this adds at most a few hundredths of a 
magnitude to the total luminosity.}. Assuming $\mu = 23.95 \pm 0.06$ and
$E(B-V) = 0.18 \pm 0.02$ leads to an absolute $g$-band magnitude $M_g = -8.9 \pm 0.2$. 
We can convert this to a standard $V$-band magnitude by adopting $M_{V,\odot} = +4.82$
and $M_{g,\odot} = +5.12$ (see the SDSS DR7 website as specified in Section 
\ref{ss:photcal}). This results in $M_V = -9.2 \pm 0.2$, rendering MGC1 a rather luminous
globular cluster with $L_{{\rm tot}} \approx 4.05 \times 10^5 L_\odot$. Our result is 
significantly more luminous than the value derived by \citet{martin:06}; however the 
difference can be entirely attributed to our larger assumed foreground extinction, as 
well as the fact that we have traced the cluster profile to a much larger radius.

Repeating this exercise for the $r$- and $i$-bands leads to apparent integrated
colours of $(g-r) = 0.64$ and $(g-i) = 1.03$. Correcting for foreground extinction and
transforming to standard magnitudes using the relations on the SDSS DR7 webpage
results in $(V-I)_0 = 0.89$ for MGC1, which is completely consistent with the integrated
colours of other remote globular clusters in both M31 \citep[e.g.,][]{huxor:09} and 
the Milky Way \citep[e.g.,][]{harris:96}. 

Finally, returning to our $g$-band profile it is straightforward to determine
the half-light radius $r_h = 2.5\arcsec \approx 7.5$ pc. This is larger than the value
of $2.3 \pm 0.2$ pc measured by \citet{martin:06}; however much of this discrepancy 
is due to our larger measured integrated luminosity -- adopting that determined
by \citet{martin:06} yields an $r_h$ smaller by at least a factor of two. A summary of
our structural and luminosity measurements for MGC1 is provided in Table \ref{t:structural}.


\section{Discussion}
The measurements we present above suggest that MGC1 lies at a distance of
$200 \pm 20$ kpc from the centre of M31. This renders it by far the most isolated
globular cluster known in the Local Group -- it resides at a much greater galactocentric
radius than do the most remote globular clusters in the Milky Way (Palomar 4 at
$R_{{\rm gc}} \sim 110$ kpc and AM 1 at $R_{{\rm gc}} \sim 120$ kpc) and the 
next most distant globular clusters known in M31 
\citep[e.g., GC10/H27 at $R_{{\rm p}} \sim 100$ kpc,][]{mackey:07,huxor:08}.
Indeed, our derived galactocentric radius for MGC1 is likely to represent a 
significant fraction of the virial radius of M31. Even so, the cluster's 
radial velocity of $\sim -50$ km s$^{-1}$ relative to M31 is consistent with that 
expected for a halo member at its observed galactocentric radius, and lies within
the probable M31 escape velocity \citep[see e.g., Fig. 4 in][]{chapman:07}. There is 
no implication, therefore, that MGC1 is anything so exotic as an unbound or intergalactic 
globular cluster. Instead, our observations suggest that the globular cluster 
populations of large spiral galaxies may be considerably more spatially extended 
than has previously been appreciated.

In this regard it is interesting that although distinct substructures have
been identified in the M31 halo to projected radii beyond $100$ kpc 
\citep[e.g.,][]{ibata:07,mcconnachie:09}, MGC1 does not lie near any known tidal stream 
or halo overdensity. Rather, it falls within an area that is apparently completely free 
of such structures -- indeed it is this region that \citet{ibata:07} use to define 
their ``clean'' M31 outer-halo stellar selection. It is also worthwhile noting
that while MGC1 lies within a $\sim 2\degr \times 2\degr$ area that contains three
satellite dwarf galaxies of M31 (And XI, XII and XIII), there is no
evidence that any of these four objects is associated with any other. On the contrary,
the three dwarfs apparently lie much closer to the M31 distance than does MGC1
\citep[][2009 in prep.]{martin:06}, while all four exhibit quite disparate radial 
velocities \citep[see e.g.,][]{chapman:07}.

\begin{table}
\centering
\caption{Structural and luminosity measurements for MGC1.}
\begin{tabular}{@{}lcc}
\hline \hline
Parameter & \hspace{2mm} & Value$^{a}$ \\
\hline
$\Sigma_g(0)$, $\Sigma_r(0)$, $\Sigma_i(0)$ & & $17.85^{{\rm m}}$, $17.17^{{\rm m}}$, $16.83^{{\rm m}}$ \\
$\Sigma_{V,0}(0)$ & & $16.95^{{\rm m}}$\vspace{1.5mm} \\
$r_c$ (King) & & $0.65\arcsec$ ($\approx 2.0$ pc) \\
$r_t$ (King) & & Indeterminate\vspace{1.5mm} \\
$r_c$ (EFF) & & $0.69\arcsec$ ($\approx 2.1$ pc) \\
$\gamma\,\,$ (EFF) & & $-2.5 \rightarrow -3.5$\vspace{1.5mm} \\
$r_h$ & & $2.5\arcsec$ ($\approx 7.5$ pc)\vspace{1.5mm} \\
$m_g$ & & $15.74^{{\rm m}}$ \\
$M_g$ & & $-8.9^{{\rm m}}$ \\
$M_V$ & & $-9.2^{{\rm m}}$ \\
$L_{{\rm tot}}$ & & $4.05 \times 10^5 L_\odot$\vspace{1.5mm} \\
$(g-r)$, $(g-i)$ & & $0.64^{{\rm m}}$, $1.03^{{\rm m}}$ \\
$(V-I)_0$ & & $0.89^{{\rm m}}$ \\
\hline
\label{t:structural}
\end{tabular}
\medskip
\vspace{-5mm}
\\
$^a$ A distance modulus of $\mu = 24.0$ and foreground extinction $E(B-V) = 0.17$ have
been used where appropriate.
\end{table}

Our measurements have also shown that MGC1 is a very luminous globular cluster.
Indeed, with $M_V \approx -9.2$ it would fit comfortably within the top ten
most luminous globular clusters in the Milky Way. The fact that it also lies at
a very large galactocentric radius reinforces the idea, first put forward by
\citet{mackey:07} \citep[see also][]{galleti:07,huxor:09}, that M31 possesses a population 
of very remote, compact, luminous globular clusters not seen in the Milky Way.
A total of seven globular clusters are known in the very outer reaches of
the Galactic halo, beyond $R_{{\rm gc}} = 40$ kpc. Of these, six are subluminous
(with $-6.0 \la M_V \la -4.7$) and rather diffuse ($11 \la r_h \la 25$ pc). 
The seventh cluster is NGC 2419, which is one of the most luminous in the
Galactic system and which has a number of unusual properties that has led to suggestions
that it may be the remains of an accreted galaxy \citep[e.g.,][]{vandenbergh:04}. 
In contrast, a search of the RBC reveals $34$ confirmed clusters in M31 with 
$R_{{\rm p}} \geq 40 \times (\frac{\pi}{4}) = 31.4$ kpc\footnote{Here the factor 
$\frac{\pi}{4}$ scales the three-dimensional radius $R_{{\rm gc}} = 40$ kpc to 
an average projected radius.}, of which $25$ are compact and $9$ are of the
extended type first reported by \citet{huxor:05}. Assuming $\mu_{{\rm M31}} = 24.47$
and a typical $E(B-V) \approx 0.1$, $12$ of the $25$ remote compact clusters
(including MGC1) have luminosities brighter than $M_V = -8.0$, while another $5$ 
have luminosities lying between this value and the peak of the GCLF at $M_V \approx -7.5$.  

Given that all but three of the $34$ remote M31 clusters come from the survey
area described in \citet{huxor:08}, covering one quadrant of the M31 halo,
this clear disparity between the remote M31 and Milky Way globular cluster systems
cannot be due simply to the global difference in population size (which would
account for, at most, a factor $\sim 3$). The origin of the differences between
the Galactic and M31 outer halo globular clusters is not known. We speculate
that it may be related to the apparently more vigorous accretion and merger
history that M31 has experienced \citep[e.g.,][]{ibata:07}; however far more detailed
study of remote M31 members will be required to shed light on this possibility.

MGC1 offers a useful probe of the outermost regions of the M31 halo, where only a
very few stars have previously been measured \citep{kalirai:06,chapman:06,chapman:08,koch:08}.
The metal abundance we derive for MGC1 is low, $[$Fe$/$H$] \approx -2.3$.
This is in contrast to results from recent high-resolution spectroscopy by
\citet{alvesbrito:09} who found $[$Fe$/$H$] = -1.37 \pm 0.15$. The origin of
this discrepancy is unclear; however we note that it also exists for the other
two clusters in their sample: GC5 and GC10, which are the next two most remote
known M31 clusters with $R_{{\rm p}} = 78$ kpc and $100$ kpc. \citet{alvesbrito:09}
derive $[$Fe$/$H$] = -1.33 \pm 0.12$ and $[$Fe$/$H$] = -1.73 \pm 0.20$, respectively, 
for these objects, while high-quality CMDs suggest $[$Fe$/$H$] = -1.84 \pm 0.15$ 
and $[$Fe$/$H$] = -2.14 \pm 0.15$ \citep{mackey:07}. Until these disagreements are 
resolved, it is too early to conclude that the mean globular cluster metallicity 
at large radii in M31 flattens off at $[$Fe$/$H$] \sim -1.6$ as suggested by
\citet{alvesbrito:09}. Taking the CMD results at face value would instead indicate
that $[$Fe$/$H$] \la -2.0$ for clusters beyond $R_{{\rm p}} \sim 80$ kpc.

The situation for M31 halo stars at radii beyond $\sim 70$ kpc is, at present, 
confusing \citep[see e.g., the discussion in][]{richardson:09}.
Spectroscopic studies by \citet{kalirai:06} and \citet{chapman:06,chapman:08} suggest 
$\left< \right. [ $Fe$/$H$] \left. \right> \sim -1.3$ in the range 
$60 \la R_{{\rm p}} \la 160$ kpc, not inconsistent with the peak of
the (photometric) metallicity distribution function (MDF) for outer-halo non-stream
stellar populations derived by \citet{ibata:07}. On the other hand, after
reanalysing of much of the spectroscopic data from the above studies, \citet{koch:08} 
found that the mean metal abundance of M31 halo stars may fall below $-2$ beyond 
$R_{{\rm p}} \sim 80$ kpc, and reach as low as 
$\left< \right. [$Fe$/$H$] \left. \right> \sim -2.5$ at $R_{{\rm p}} \sim 160$ kpc.
The MDF of \citet{ibata:07} appears to exhibit a metal-poor tail and
our results for MGC1 suggest that there must be at least {\it some} very metal-poor
component to the extremely remote outer halo of M31. However on the whole
it seems that the most distant globular clusters in M31 are more metal-poor
than the bulk of the field halo stars at comparable radii.

Another remarkable feature of MGC1 is its extremely extended outer structure.
Cluster members are clearly visible at large radii, and these are apparently
fairly homogeneously distributed azimuthally. We have traced the radial profile 
of MGC1 to at least $\sim 450$ pc and possibly as
far as $\sim 900$ pc, rendering it by far the most extended globular cluster
hitherto studied. The most extended cluster in the Milky Way is NGC 2419,
with $r_t \approx 200$ pc; the vast majority of the Galactic population have
$r_t \la 100$ pc \citep[e.g.,][]{mackey:05}. The unique structure of MGC1
likely reflects its unmatched remote location, away from any strong tidal forces
which might act to impose a tight limit on its radial extent. The expected tidal radius
of a globular cluster of mass $M_{{\rm cl}}$ orbiting a galaxy of mass $M_{{\rm g}}$
at a galactocentric radius $R_{{\rm gc}}$ is given by:
\begin{equation}
r_t = R_{{\rm gc}} \left( \frac{M_{{\rm cl}}}{3 M_{{\rm g}}} \right)^{\frac{1}{3}} \,\,.
\end{equation}
Adopting our measured parameters for MGC1 ($R_{{\rm gc}} = 200$ kpc and
$L_{{\rm tot}} = 4.05 \times 10^5 L_\odot$) along with a typical $V$-band
mass-to-light ratio for a globular cluster $M / L \approx 2$ and the mass
of M31 from \citet{evans:00}, $M_{{\rm g}} \approx 1.2 \times 10^{12} M_\odot$,
we find the expected tidal radius for MGC1 is $r_t \sim 1200$ pc.
It is therefore not surprising we have not identified a tidal cut-off in
its radial profile.

Note that we cannot be certain the outermost cluster members are bound to the
cluster, since any tidal distortions would not necessarily lie in the plane of the sky.
However, the lack of any clear spatially asymmetric distribution of these outer
members together with the calculation above showing a very large expected $r_t$
suggests more of a halo-type structure than strong tidal tails.

Assuming a bound halo of stars around the cluster, the observed power-law fall-off of 
the profile is not unexpected.
\citet*{baumgardt:02} used $N$-body simulations to study the long-term evolution of 
isolated star clusters. Their models show that an isolated cluster tends to
build up a surrounding halo of stars that have been scattered onto radial orbits
by two- or three-body encounters in the inner regions of the cluster.
The resulting density profile follows a power-law fall-off as $\rho \sim r^{-3.3}$
at large radii. 
This is very similar to the results from, for example, the Monte Carlo models of
\citet{spitzer:87} which predict $\rho \sim r^{-3.5}$. In projection we thus
expect a power-law fall-off with an exponent somewhere between $-2.5 \la \gamma \la -2.3$, 
which is entirely consistent with what we observe for the inner profile of MGC1. 

In both sets of models the core-halo structure takes many median relaxation
times to establish over radial ranges of up to $\sim 100 r_h \approx 750$ pc
for MGC1. It is therefore perhaps not surprising that we observe a break
to a steeper fall-off in the outer regions of MGC1. This may reflect the
ongoing initial build-up of the cluster's halo, or possibly the re-establishment of 
this halo after the last pericentre passage 
of MGC1 which could have brought it much closer to M31 than it is at present.
Whatever the case, given that the median relaxation time of MGC1 is 
likely to be several Gyr 
\citep[see, for example, the values listed for Galactic globulars by][]{harris:96},
the existence of a core-halo structure at all suggests that MGC1 has been evolving 
in isolation for a very considerable period of time.

\section{Conclusions}
In summary, we have presented resolved photometric measurements of MGC1 from deep 
Gemini/GMOS imaging of this remote M31 cluster. Our resulting CMD displays features 
consistent with those of an ancient stellar population, including a long horizontal 
branch populated across the instability strip. By fitting isochrones calculated by 
several different groups, as well as fiducial sequences observed for Galactic
globular clusters, we determine MGC1 to be a metal-poor object with
$[$M$/$H$] \approx -2.3$. This is in disagreement with
a recent high resolution spectroscopic measurement by \citet{alvesbrito:09} who find
a significantly higher metal abundance; the origin of this discrepancy is not yet
clear but it also exists for several other remote M31 globular clusters.

The best-fitting stellar evolution models and globular cluster fiducials suggest that 
MGC1 lies significantly closer 
to us than M31. Our preferred distance modulus is $\mu = 23.95 \pm 0.06$ (corresponding
to a line-of-sight distance of $\approx 620$ kpc), compared with the generally accepted 
value for M31 of $\mu_{{\rm M31}} = 24.47$ ($\approx 780$ kpc). Combined with 
the large projected separation between MGC1 and M31 ($R_{{\rm p}} = 117$ kpc), 
our distance measurement implies that MGC1 lies in the extreme outer regions of the 
M31 halo with a galactocentric radius $R_{{\rm gc}} = 200 \pm 20$ kpc. This renders
it the most isolated known globular cluster in the Local Group by some considerable
margin. Even so, the radial velocity measured for this object 
\citep{galleti:07,alvesbrito:09} is within the M31 escape velocity and 
consistent with that expected for an M31 halo member at the radius of MGC1.

By measuring the radial brightness profile of MGC1 we have established that this
cluster is centrally concentrated with a core radius $r_c \approx 2$ pc. It is also
rather luminous with $M_V = -9.2$. This reinforces the notion, first put forward by
\citet{mackey:07}, that M31 possesses a population of compact, luminous globular 
clusters in its outer halo that does not have a counterpart in the Milky Way.

Perhaps unsurprisingly given its extremely remote location, MGC1 does not show
any evidence for a tidal limit in its outer regions, and King models do not provide
a good description of the shape of the brightness profile at large radii. Instead we 
observe a power-law fall-off with exponent $\gamma = -2.5$, breaking to $\gamma = -3.5$
in the outermost parts. This core-halo structure, which we see extending to radii of 
at least $450$ pc and possibly as far as $\sim 900$ pc, is broadly in line with 
expectations derived from numerical modelling of isolated globular clusters.
In this regard MGC1 is unique in our experience -- the most radially extended globular
cluster in the Milky Way is NGC 2419 which shows a tidal cut-off at $r_t \approx 200$ pc.
Further detailed observational study of MGC1 may therefore be extremely useful for 
advancing our understanding of globular cluster evolution.

\section*{Acknowledgments}
ADM and AMNF are supported by a Marie Curie Excellence Grant from the European 
Commission under contract MCEXT-CT-2005-025869. NRT acknowledges financial
support via a STFC Senior Research Fellowship.
This work is based on observations obtained at the Gemini Observatory, which is operated 
by the Association of Universities for Research in Astronomy, Inc., under a cooperative 
agreement with the NSF on behalf of the Gemini partnership: the National Science Foundation 
(United States), the Science and Technology Facilities Council (United Kingdom), the
National Research Council (Canada), CONICYT (Chile), the Australian Research Council
(Australia), Minist\'{e}rio da Ci\^{e}ncia e Tecnologia (Brazil) and SECYT (Argentina).
These observations were obtained under program GN-2007B-Q-58.

\bsp
\label{lastpage}

\end{document}